\documentclass[prd,aps,twocolumn,a4paper,showkeys,nofootinbib]{revtex4-1}

\usepackage{graphicx,psfrag}
\usepackage{mathrsfs}
\usepackage{amsmath,amsfonts,amssymb}
\usepackage{multirow}
\usepackage{comment}
\usepackage{ulem}
\usepackage{hyperref}
\usepackage{enumitem}
\usepackage{tcolorbox}
\usepackage{xspace}
\usepackage{enumerate}

\usepackage{xcolor}

\newcommand{\be}{\begin{equation}}
\newcommand{\ee}{\end{equation}}
\newcommand{\bea}{\begin{eqnarray}}
\newcommand{\eea}{\end{eqnarray}}
\newcommand{\bel}{\begin{align}}
\newcommand{\eel}{\end{align}}

\newcommand{\tGRAthena}{\texttt{GR-Athena++}}
\newcommand{\GRAthena}{\tGRAthena\xspace}

\newcommand{\tAthena}{\texttt{Athena++}}
\newcommand{\Athena}{\tAthena\xspace}

\newcommand{\tLorene}{\texttt{Lorene}}
\newcommand{\Lorene}{\tLorene\xspace}

\newcommand{\POLUU}{\texttt{POL-UU}}
\newcommand{\HRPOLUU}{\texttt{HR-POL-UU}}
\newcommand{\POLUD}{\texttt{POL-UD}}
\newcommand{\TORPP}{\texttt{TOR-PP}}
\newcommand{\TORPM}{\texttt{TOR-PM}}
\newcommand{\HRMIX}{\texttt{HR-MIX}}
\newcommand{\MIX}{\texttt{MIX}}
\newcommand{\MIXED}{\texttt{MIX}}
\newcommand{\NOB}{\texttt{B0}}
\newcommand{\LOWB}{\texttt{LOWB}}
\newcommand{\POLLR}{\texttt{POL-90D}}
\newcommand{\POLASYM}{\texttt{POL-ASYM}}
\newcommand{\STIFF}{\texttt{STIFF}}
\newcommand{\BIT}{\texttt{BITANT}}
\newcommand{\BITANT}{\texttt{BITANT}}
\newcommand{\POLQ}{\texttt{POL-Q12}}

\def\gccm{{\rm g\,cm^{-3}}}
\def\Msun{{\rm M_{\odot}}}

\def\GMc2{{\rm G M_{\odot} c^{-2}}}

\def\kt2{\kappa^\text{T}_2}

\def\kt2{\kappa^\text{T}_2}

\def\2nd{2^\mathrm{nd}}
\def\4th{4^\mathrm{th}}
\def\6th{6^\mathrm{th}}
\def\8th{8^\mathrm{th}}

\def\z4c{$\mathrm{Z}4\mathrm{c}$}
\def\z4oc{$\mathrm{Z}4(\mathrm{c})$}
\def\z4{$\mathrm{Z}4$}
\def\ccz4{$\mathrm{CCZ}4$}

\def\ergsec{{\rm erg\,s^{-1}}}
\def\gccm{{\rm g\,cm^{-3}}}
\def\Msun{{\rm M_{\odot}}}

\def\GMc2{{\rm G M_{\odot} c^{-2}}}

\usepackage{pifont} 

\usepackage{color}
\definecolor{cyan}{rgb}{0,0.9,0.9}
\definecolor{orange}{rgb}{0.9,0.5,0}
\definecolor{magenta}{rgb}{1,0,1}
\definecolor{purple}{rgb}{0.8,0.4,0.8}
\definecolor{gray}{rgb}{0.8242,0.8242,0.8242}
\definecolor{light-gray}{gray}{0.95}

\begin{document}

\title{Magnetic Field Configurations in Binary Neutron Star Mergers II: Inspiral, Merger and Ejecta}

\author{William Cook$^1$}
\author{Eduardo M. Guti\'errez$^{2,3}$}
\author{Sebastiano Bernuzzi$^1$}
\author{David Radice$^{2,3,4}$}
\author{Boris Daszuta$^{1}$}
\author{Jacob Fields$^{2,3}$}
\author{Peter Hammond$^{2,3,5}$}
\author{Harshraj Bandyopadhyay$^3$}
\author{Maximilian Jacobi$^1$}

\affiliation{%
$^1$Theoretisch-Physikalisches Institut, Friedrich-Schiller-Universität Jena, 07743, Jena, Germany \\
$^2$Institute for Gravitation and the Cosmos, The Pennsylvania State University, University Park, Pennsylvania, 16802, USA \\
$^3$Department of Physics, The Pennsylvania State University, University Park, Pennsylvania, 16802, USA \\
$^4$Department of Astronomy and Astrophysics, The Pennsylvania State University, University Park, Pennsylvania, 16802, USA \\
$^5$Department of Physics and Astronomy, University of New Hampshire, Durham, NH 03824, USA
}

\date{\today}

\begin{abstract}
We perform a series of simulations of magnetised Binary Neutron Star mergers, with varying magnetic field topologies in the initial data, as well as varying Equations of State, and mass ratios.
In this paper, a companion paper to \cite{Gutierrez:2025gkx}, we analyse the impact of the initial field configuration on the emitted gravitational wave signal, the global amplification of the magnetic field, 
and the nature of the ejected material from the binary. We investigate the dependence of the phase evolution of the gravitational wave signal in the post-merger on the initial magnetic field, finding that dephasing effects between the dominant $(\ell=2,m=2)$ mode of the gravitational wave signal, and leading subdominant modes $(2,1),(3,3)$ may be strongly impacted by the choice of numerical reconstruction scheme. The 
magnetic field amplification at merger may be considerably enhanced during the Kelvin-Helmholtz instability dominated phase
by the presence of anti-aligned fields between the stars, or suppressed by the presence of toroidal fields; while the post-merger 
amplification of the field due to winding may also be suppressed by toroidal fields, and may be enhanced by asymmetries or mixtures of poloidal and toroidal fields. 
The magnetic field strength in the ejecta may also be strongly impacted by the nature of the initial magnetic 
field configuration, with initial data configurations which lead to large amplifications and those with mixtures of poloidal and toroidal fields preferentially emitting 
highly magnetised material in the polar regions, while we find that the magnetic field strength in the ejecta for these configurations shows a weaker dependence on the density of the material than in those cases that amplify the magnetic field less. We investigate the orientation of the field in the ejecta, and find that 
the magnetic field is largely randomly oriented in the ejected material, providing supporting evidence for such models used to estimate thermalisation timescales of ejected material.
Further we find that configurations which begin with an initial bitant symmetry break this symmetry in a uniform manner, independent of the initial magnetic field configuration, when evolved without an
enforced symmetry. This behaviour suggests the presence of a spontaneous symmetry breaking bifurcation in the solution.
\end{abstract}


\maketitle

\section{Introduction}
\label{sec:intro}

Neutron stars (NSs) can be the site of some of the strongest magnetic fields in the universe, 
reaching strengths of up to $10^{16} {\rm G}$. Consequently, magnetic fields can play a key role
in the evolution of binary neutron star (BNS) systems, both during the moment of merger,
and in the post-merger phase. At the moment of first contact between two NSs,
it is expected that a thin shearing layer will form, within which the Kelvin-Helmholtz instability
(KHI) will be triggered, forming small vortex like structures in the fluid, 
which wind the magnetic field with them, amplifying the magnetic field energy by at least 6
orders of magnitude \citep{Rasio:1999a,Price:2006fi,Kiuchi:2015sga}. Strong magnetic fields can lead to additional pressure and tension forces
in the fluid 
evolution which can drive the ejection of neutron rich material during a BNS merger \citep{Siegel:2014ita,Kiuchi:2014hja,Siegel:2017nub,Mosta:2020hlh,Curtis:2021guz,Combi:2022nhg,deHaas:2022ytm,Kiuchi:2022nin,Combi:2023yav,Most:2025kqf}, 
material which subsequently will undergo r-process nucleosynthesis \citep{Pian:2017gtc,Kasen:2017sxr}, leading to the formation of
heavy elements, visible in a kilonova \citep{Li:1998bw,Kulkarni:2005jw,Metzger:2010sy,GBM:2017lvd,Coulter:2017wya,Soares-santos:2017lru,Arcavi:2017xiz}.

Strong magnetic fields in the post-merger remnant have been proposed as a mechanism 
for the launching of ultra-relativistic jets \citep{Blinnikov:1984a,Paczynski:1986px,Goodman:1986a,Eichler:1989ve,Narayan:1992iy}, which are believed to lead to the production of 
gamma ray bursts 
such as GRB170817, seen coincidentally with the gravitational wave (GW)
signal from the BNS merger GW170817 \citep{TheLIGOScientific:2017qsa,Goldstein:2017mmi,Savchenko:2017ffs}. Such jets may be launched either from a highly magnetised remnant NS,
a so-called magnetar, or from a magnetised black hole (BH) with an associated accretion disk, through the 
Blandford-Znajek process, resulting from gravitational collapse after the BNS merger \citep{Piran:2004ba,Kumar:2014upa,Ciolfi:2018tal}

Observations of pulsars suggest a large scale dipolar field \citep{Chung:2011a,Chung:2011b,Bilous:2019,Riley:2021pdl,Choudhury:2024,Salmi:2024}
with contributions from higher multipoles,
as the dominant exterior contribution to the magnetic field in a neutron star, arising both during the formation of the star; and through convective and thermal processes occurring in the early star evolution \cite{Pons:2019zyc}.
The details of the interior magnetic field of the star are however less well constrained. 
Long term simulations of static neutron stars with initially purely poloidal fields \citep{Kiuchi:2008ss,Ciolfi:2011xa,Ciolfi:2013dta,Lasky:2011un,Pili:2014npa,Pili:2017yxd,Sur:2021awe,Cook:2023bag,Cook:2025zzy} suggest that well known magnetohydrodynamical (MHD) instabilities
associated to $m=0$ and  $m=1$ mode instabilities of poloidal configurations are important in the initial rearrangement of the field, developing a non-trivial toroidal component on 
Alfven timescales \citep{Tayler:1957a,Tayler:1973a,Wright:1973a,Markey:1973a,Markey:1974a,Flowers:1977a,Braithwaite:2005md}.
Magnetised BNS simulations are, however, frequently initialised with purely poloidal initial field profiles, with recent exceptions 
\cite{Kawamura:2016nmk,Ruiz:2020via,Aguilera-Miret:2024cor,Bamber:2024wqr,Tsokaros:2024wgb}. For binary systems which have evolved over many Alfven timescales, we can view the constituent stars as effectively isolated from their binary partner from 
the perspective of their electromagnetic and thermodynamical evolution, before they approach sufficiently close through GW emission that the magnetospheres of the stars intersect, or for tidal 
effects to become appreciable in magnitude.  During this effectively isolated evolution, we can expect that such initial linear instabilities, which operate over Alfven timescales
 have had sufficient time to saturate and that a full non-linear rearrangement of the magnetic field has occurred. 
Equilibrium configurations of magnetic fields inside neutron stars have been constructed 
with incredibly strong toroidal magnetic fields \citep{Ciolfi:2013dta}, 
and some numerical studies have suggested that the presence of an initially 
strong toroidal magnetic field may be preserved over multiple Alfven
timescales \citep{Sur:2020hwn,Sur:2021awe}. If such configurations are
indeed dynamically stable, they may be found in the interior of stars in 
binary systems.

Recent works have investigated the impact of magnetic field topology on BNS evolutions.
In \cite{Bamber:2024wqr,Tsokaros:2024wgb} the impact on the post-merger gravitational wave signal has been investigated,
finding that the configuration of the magnetic field may have a detectable impact on the peak post-merger frequency.
Magnetic field configurations have also been compared in \cite{Aguilera-Miret:2021fre,Aguilera-Miret:2023qih,Aguilera-Miret:2024cor},
with results suggesting that the end state of the merger process results in a universal magnetic field configuration, independent of
the initial data. Contrastingly, simulations performed in \cite{Chabanov:2022twz} have observed different magnetic field amplifications
for a magnetic field localised in the crustal region of the star. Beyond magnetic field topologies, the results of \cite{Aguilera-Miret:2025nts}
have suggested further universality in the extent of the amplification of magnetic fields, independent of the mass ratios, and Equations of State (EOS) considered. 
In \cite{Kawamura:2016nmk} the impact of the magnetic field orientation on jet launching is investigated, finding that the formation
of a large scale field is independent of the initial condition, while \cite{Ruiz:2020via} find that the jet launching time, and ejecta, may
be very sensitive to the specific magnetic field configuration.

The nature of the magnetic field within the ejected material from the binary may play an important role in understanding 
its evolution, and the associated kilonova. It has been suggested that strong magnetic fields in the ejected matter can affect the r-process nucleosynthesis, through the quantisation of 
electron momentum states into Landau levels, affecting $\beta-$ and neutron decay rates in the ejected material \citep{Famiano:2020fbq,Tambe:2024usx,Kumamoto:2024jiq} .
In addition the nature of the magnetic field configuration in the ejected material may affect the ability of the ejected material to fully thermalise,
having a strong impact on the luminosity of kilonova light curves \citep{Barnes:2016umi}.

This paper serves as a companion paper to \cite{Gutierrez:2025gkx} (henceforth Paper I). We perform and discuss the 
results of a sequence of simulations, investigating different magnetic field configurations, different
symmetry considerations, different Equations of State, and different mass ratios. In particular, we perform 
simulations of poloidal magnetic field configurations with different geometries that are misaligned with each 
other to investigate generic configurations. We also perform  simulations of magnetic field configurations with 
a combination of poloidal and toroidal fields which aims to reproduce the effect of the saturation of the 
initial instabilities of a purely poloidal magnetic field profile, approximating configurations found in  \cite{Cook:2025zzy}.  
In Paper I we focus on the resultant evolution of the post-merger remnant and disk. In this paper we focus on the 
gravitational wave signal, global magnetic field dynamics at merger, and the evolution of the ejected matter. 
We perform simulations in this paper using the numerical relativity (NR) code 
\GRAthena \citep{Daszuta:2021ecf,Cook:2023bag,Daszuta:2024chz,Daszuta:2024ucu}, based on the \Athena infrastructure \citep{Stone:2020}.

In Section \ref{sec:meth} we discuss the initial data configurations we consider, and the numerical approach taken in our simulations.
In Section \ref{sec:res} we discuss the results of our simulations, focussing on the GWs (Sec. \ref{ssec:gw}), 
the magnetic field amplification (Sec. \ref{ssec:bamp}),  the symmetry of the 
configurations found (Sec. \ref{sec:sym}) and the ejected material from the binary (Sec. \ref{ssec:ej}). We summarise our findings in Sec. \ref{sec:con}. In Appendix \ref{app:recon}, we discuss the errors 
associated to the numerical reconstruction procedure present in our results. Unless stated otherwise, our results are presented in CGS units.

\section{Method}
\label{sec:meth}

We perform a series of magnetised binary merger simulations using the \GRAthena code,
with varying initial magnetic field configurations. We refer readers to Table 1
in Paper I, which details the overall parameters of the configurations considered, and below recap the key details of the simulations performed.

\subsection{Numerical details}

\GRAthena employs task based parallelism and an oct-tree, block based grid structure, to ensure 
highly scaling performance on large numbers of CPU cores, as described in \cite{Stone:2020}. 
The Einstein equations are evolved using 
the Z4c formulation \citep{Bernuzzi:2009ex,Ruiz:2010qj,Weyhausen:2011cg,Hilditch:2012fp}, with the 
moving puncture gauge \citep{Brandt:1997tf,Campanelli:2005dd,Baker:2005vv}, using a vertex centred 
grid and 6th order finite differencing stencils. For discussion of the vertex centred implementation
we refer the reader to \cite{Daszuta:2021ecf,Daszuta:2024ucu}. 
We refer the reader to \cite{Cook:2023bag} for the choice of gauge, constraint damping and 
dissipation parameters, which remain identical between all runs considered in this paper 
and the BNS simulations presented there. 

The relativistic Euler equations are evolved in the standard 3+1 ``Valencia'' formulation, with 
the Maxwell equations included in the ideal MHD approximation \citep{Anile:1990a,Marti:1991wi,
Banyuls:1997zz,Balsara:1998b,Komissarov:1999a,Gammie:2003rj,Anninos:2005kc,Komissarov:2005wj,
Anton:2005gi,DelZanna:2007pk}.
Evolution of the relativistic Euler equations is performed using a Godunov style finite volume approach, with the Local Lax Friedrichs (LLF) approximate Riemann solver used to calculate a numerical flux function, and 
primitives reconstructed to the cell interface with either the PPM \citep{Colella:2011a,McCorquodale:2015a} or WENOZ \citep{Borges:2008a} stencil for high resolution shock capturing (see Appendix \ref{app:recon}). In addition to the Euler equations we evolve the 
electron fraction $Y_e$ as a scalar advected with the fluid density as described in \cite{Cook:2023bag}. The equation of state is given by tabulated data for the SFHo \cite{Steiner:2012rk} and DD2 \cite{Typel:2009sy} equations of state from the Compose database \citep{Typel:2013rza}. 
External to the evolved stars we set a low density
unphysical atmosphere to avoid excessively sharp gradients at the star surface, and to avoid the breakdown of the weak solutions to the Euler equations in vacuum. The density of this atmosphere is set to $\rho_{\rm atm} = 1.8\times 10^3 {\rm g/cm}^3$ and atmosphere temperature to 0.1 MeV. The electron fraction of the atmosphere is 0.5.
All simulations are performed with a standard resolution (SR) of 180m on the finest refinement level fully covering the stars,
other than two, shorter lived, high resolution (HR) configurations with a finest resolution of 123m. 

The magnetic field is evolved using the constrained transport algorithm implementation of \cite{Gardiner:2007nc,White:2015omx}, with the divergence and curl preserving restriction and prolongation operators of \cite{Toth:2002a},  utilising a face averaged sampling of the magnetic field. This provides conservation of the magnetic field divergence to machine precision, as demonstrated in \cite{Cook:2023bag}. Time evolution is performed using a third order Runge-Kutta time integrator with CFL factor 0.25. 
Conserved-to-Primitive variable inversion is implemented through a custom solver ``PrimitiveSolver'', detailed in Appendix A of \cite{Cook:2023bag}, based on the RePrimAnd algorithm of \cite{Kastaun:2020uxr}.
We extract gravitational waves using the Newman-Penrose scalar $\Psi_4$ as sampled on geodesic spheres, to avoid coordinate singularities and sampling issues near the poles of extraction spheres sampled equally in standard $(\theta,\phi)$ spherical coordinates.

\subsection{Hydrodynamical Initial data}
\label{sec:HDID}
For the simulations considered in this paper, we generate initial data with the \Lorene 
initial data library \citep{Gourgoulhon:2000nn}. We consider 3 different hydrodynamical configurations. First, an equal mass model with SFHo EOS, corresponding to a binary with ADM mass $M_{\rm ADM} = 2.672 \Msun$, initial orbital frequency $331.275$Hz, orbital angular momentum $J = 7.001 \Msun/c^2$,
initial separation 40.0 km, baryon mass $M_{\rm b} = 2.967 \Msun$ and gravitational mass $2.7004 \Msun$. 
We perform one simulation with the SFHo EOS with mass ratio $q=1.2$, with ADM mass $M_{\rm ADM} = 2.675 \Msun$, initial orbital frequency $281.407$Hz, orbital angular momentum $J = 7.212 \Msun/c^2$,
initial separation 45.0 km, total baryon mass $M_{\rm b} = 2.970 \Msun$ and gravitational mass $2.7002 \Msun$. 
We also perform one simulation with the DD2 EOS, with ADM mass $M_{\rm ADM} = 2.672 \Msun$, initial orbital frequency $331.567$Hz, orbital angular momentum $J = 7.001 \Msun/c^2$,
initial separation 40.0 km, baryon mass $M_{\rm b} = 2.941 \Msun$ and gravitational mass $2.7001 \Msun$. 
All configurations are initialised with temperature $T=0.1\ \mathrm{MeV}$ and assuming
cold $\beta$-equilibrium.

\subsection{Magnetic Field Configurations}
\label{sec:bconfig}

\begin{table*}[t]
  \centering    
  \begin{tabular}{c|ccccc}        
    \hline
    Run Name & $B_0$ &Resolution& EOS & Recon. & Field configuration \\
    \hline
    \POLUU& $5\times 10^{15} \mathrm{G}$ &SR & SFHo & WENOZ & Aligned Poloidal \\
    \HRPOLUU& $5\times 10^{15} \mathrm{G}$ &HR  & SFHo & WENOZ & Aligned Poloidal  \\
    \MIXED& $5\times 10^{15} \mathrm{G}$ &SR  & SFHo & WENOZ & Superposed Poloidal and Toroidal  \\
    \HRMIX& $5\times 10^{15} \mathrm{G}$ &HR  & SFHo & WENOZ & Superposed Poloidal and Toroidal  \\
    \TORPP& $5\times 10^{15} \mathrm{G}$ &SR  & SFHo & WENOZ & Aligned Toroidal  \\
    \TORPM& $5\times 10^{15} \mathrm{G}$ &SR  & SFHo & WENOZ & Anti-aligned Toroidal  \\
    \NOB& $0 \mathrm{G}$ &SR  & SFHo & WENOZ & No Field \\
    \BITANT& $5\times 10^{15} \mathrm{G}$ &SR  & SFHo & PPM & Aligned Poloidal with Bitant Sym.  \\
    \LOWB& $5\times 10^{8} \mathrm{G}$ &SR  & SFHo & PPM & Aligned Poloidal  \\
    \POLUD& $5\times 10^{15} \mathrm{G}$ &SR  & SFHo & PPM & Anti-aligned Poloidal  \\
    \POLLR& $5\times 10^{15} \mathrm{G}$ &SR  & SFHo & PPM & Poloidal in orbital plane \\
    \POLASYM& $5\times 10^{15} \mathrm{G}$ &SR  & SFHo & PPM & Displaced Aligned Poloidal \\
    \STIFF& $5\times 10^{15} \mathrm{G}$ &SR  & DD2 & PPM & Aligned Poloidal  \\
    \POLQ& $5\times 10^{15} \mathrm{G}$ &SR  & SFHo & PPM & Aligned Poloidal, Mass ratio $q=1.2$ \\
    \hline
  \end{tabular}
  \caption{Summary of runs performed. $B_0$ represents the maximum magnetic field strength. Recon. describes the reconstruction scheme used.}
 \label{tab:runs}
\end{table*}

In order to investigate the impact of toroidal fields which may develop as the result of instabilities in purely poloidal fields as discussed above, we perform a sequence of runs with purely poloidal and
purely toroidal magnetic fields, as well as a run with superposed poloidal and toroidal fields, as a model of the 
configuration that can be expected to develop in an isolated star over many Alfven times, which we can expect to 
have developed by the time that a binary system reaches its final few orbits.

In order to construct a purely poloidal magnetic field we initialise the field 
through taking the curl of a purely toroidal vector potential defined as,

\bea
A^\phi &=& B_0 \max(p - p_{\mathrm{cut}},0)^{n_s} \\
(A^x,A^y A^z) &=&  (-yA^\phi,xA^\phi,0),  \label{eq:poloidalA}
\eea
with parameters set as detailed in Paper I. In particular $B_0$ is set such that the magnetic field strength is $5\times 10^{15}$G. 
We refer to this configuration as \POLUU.
To investigate the impact of a toroidal field we use the vector potential specified in Eq. \ref{eq:toroidalA}, similar to that used in \cite{Hayashi:2022cdq,Bamber:2024wqr},

\bea
{\bf A} &=& B_0 \max(p - p_{\mathrm{cut}},0)^{n_s}\nonumber\\&&(x(z^2-R^2),y(z^2-R^2),-z(x^2+y^2-R^2)). \label{eq:toroidalA}
\eea

For the initial data considered in Sec. \ref{sec:HDID}, the stars are initially non-spinning.
In long term simulations of magnetised non-spinning stars, there is no process expected to 
globally break the rotational symmetry, and create a toroidal field with a globally coherent orientation.
This is in contrast to a rotating star, whose angular momentum could be expected to break symmetry and 
give a global orientation to the toroidal field profile. The choice of toroidal initial data given in 
Eq. \ref{eq:toroidalA}, is antisymmetric over the $z$ axis, so has no overall favoured orientation
within the star itself. The binary system, in contrast, has an initial orbital angular momentum, 
picking out a global orientation which the toroidal field in the remnant can be expected to 
adopt. We therefore perform two runs with purely toroidal field orientations, to assess the impact of 
the relative orientation of the toroidal field and the orbital angular momentum. The first run, denoted
\TORPP~ has the parameter $B_0$ positive in both stars, while the configuration
\TORPM~ sets the parameter $B_0$ positive in the first star, and negative in the second star.

Finally we consider the orientation \MIX.
For this magnetic field configuration we superpose the \POLUU~ and \TORPP~
configurations, scaling the initial strength parameters $B_0$ to obtain a relative 
field strength of $~85\%$ poloidal, to $15\%$ toroidal, approximately matching the 
field strengths obtained in single star simulations after the saturation of the 
initial varicose and kink instabilities after the initial few Alfven times, such as those seen in \cite{Cook:2025zzy}. 
We do not expect this configuration to be an exact match for the field rearrangement at
late times in an isolated single star, but to give some insight into the impact of a mixed
poloidal and toroidal field in a binary merger.

A generic binary system containing poloidal fields may be expected to contain fields not aligned with the orbital angular momentum. To investigate the impact of moving beyond a symmetric magnetic field configuration we consider modifications of the poloidal configuration also.
We consider a configuration \POLUD, with the same magnetic field configuration as \POLUU, with the orientation of the field in one star flipped with respect to the $z$ axis.
The configuration \POLLR~ has both poloidal magnetic fields rotated to lie in the orbital plane ($z=0$), initially pointing towards each other along their line of sight. 
The configuration \POLASYM~ takes the configuration of \POLUU, and shifts the field by 147m in the positive $z$ direction, thus breaking the symmetry of the configuration.
Configurations \POLQ, \STIFF, \LOWB~ and \BIT~ all have the same magnetic field configuration as \POLUU, but differ due to a mass ratio $q=1.2$, a DD2 EOS, a low initial field strength $B_0 = 5\times 10^8 G$ and an enforced bitant symmetry respectively. Other than \BITANT, all runs have no imposed grid symmetry. We also evolve a reference configuration \NOB~ with no magnetic field.

\section{Results}
\label{sec:res}
\subsection{Gravitational Waves}
\label{ssec:gw}
We first present the gravitational waveforms from the performed simulations. 
We present the gravitational wave strain obtained through the double time integration of the Newman-Penrose scalar $\Psi_4$, directly
extracted from the numerical simulation on spheres of radius $r_{\mathrm{ex}} = 400 M_\odot \approx 588$ km centred around the centre of mass of the system.
The integration is performed using the fixed frequency integration (FFI) of \cite{Reisswig:2010di}. We define the GW strain $h$, amplitude $A$, phase $\phi$ 
and instantaneous frequency $\omega$, as decomposed onto spherical harmonics of spin weight $-2$; and 
retarded time coordinate $u$,

\begin{eqnarray}
\Psi_{4,(\ell,m)} &=& \ddot{h}_{\ell,m},\\
r_{\mathrm{ex}}h_{\ell,m} &=& A_{\ell,m}e^{-i\phi_{\ell,m}},\\
\omega_{\ell,m} &=& \dot{\phi}_{\ell,m},\\
u &=& t - R - 2M\log\left(\frac{R}{2M} - 1 \right),\\
R &=& r_{\mathrm{ex}}\left(1 + \frac{M}{2r_{\mathrm{ex}}}\right)^2,
\end{eqnarray}
where $M$ is the gravitational mass.

\begin{figure*}[t]
  \centering
    \includegraphics[width=0.99\textwidth]{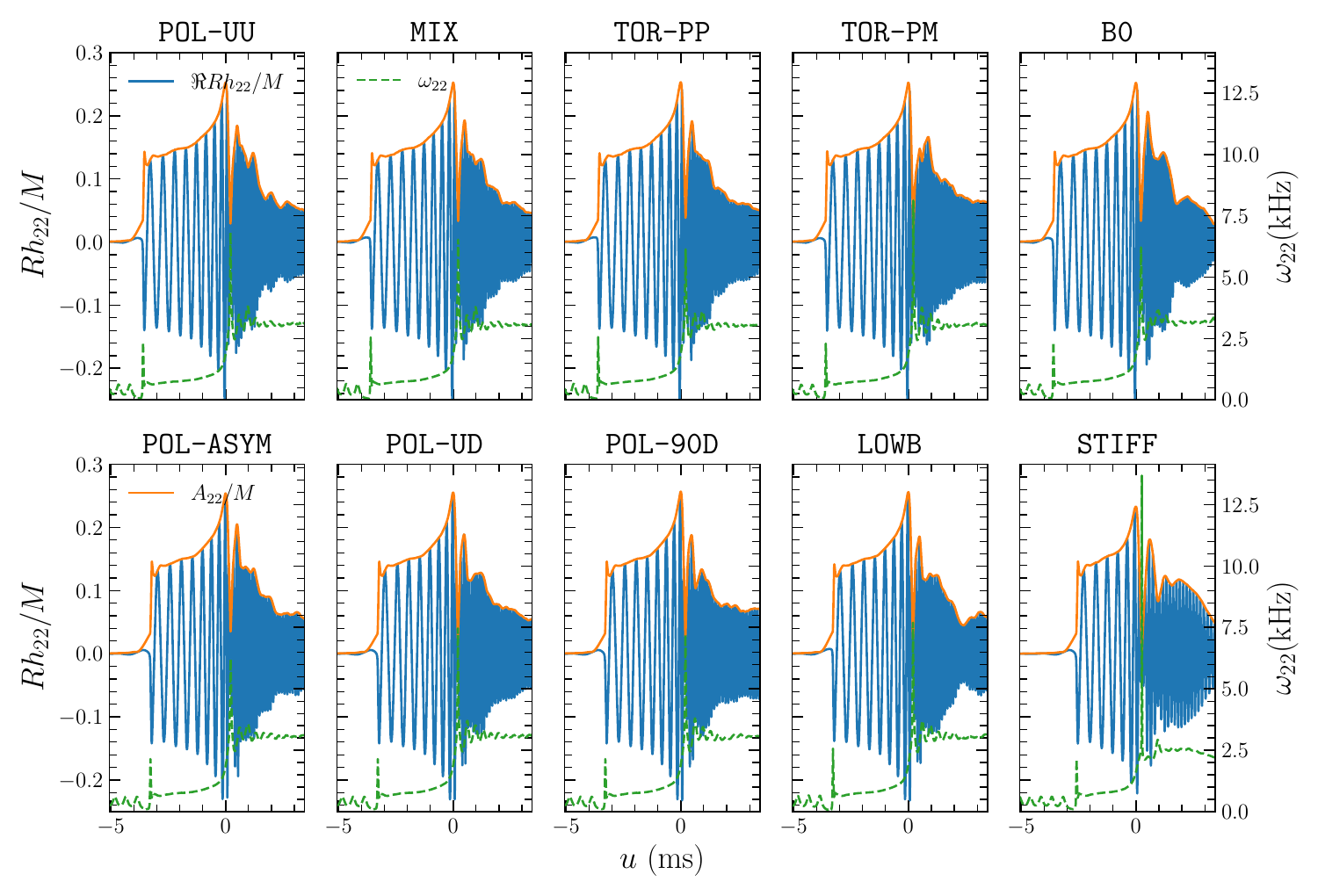}
    \caption{The real part of the $(\ell,m) = (2,2)$ mode of the gravitational wave strain $h_{22}$ and its overall amplitude (left axes) and its instantaneous frequency (right axes); extracted at a radius of 588 km. 
}
 \label{fig:GW}
\end{figure*}

We demonstrate the GW strains for the equal mass configurations in Fig. \ref{fig:GW}.
Due to the comparatively small energy density in the magnetic field
compared to the rest mass energy of the NSs, the inspiral phase is almost identical between the considered simulations, with a close match in the 
GW signals during this phase. 
All equal mass simulations complete the same number of orbits (3),
irrespective of any difference in hydrodynamical evolution.
In the post-merger phase however, we see some variations in the GW signal for different magnetic field configurations, for different GW modes,
as also seen by \cite{Bamber:2024wqr,Tsokaros:2024wgb}.
For the dominant $(\ell,m) = (2,2)$ mode, we see that the post-merger behaviour is broadly similar between all configurations, with the gravitational wave frequency suddenly increasing post-merger as the overall amplitude decays. While most configurations show a clear secondary peak in the GW amplitude post-merger, the run \TORPM, 
demonstrates an initial subdominant peak before falling and rising again to a second peak, which is the largest amplitude 
for this configuration post-merger.  The instantaneous frequency evolution of the post-merger GW is relatively consistent between 
configurations with different magnetic field geometries and different magnetic field strengths. In all SFHo cases, we see a peak of the 
gravitational wave frequency at merger before the frequency quickly reduces to $3.06\mathrm{kHz} $, while 
for the \STIFF~ EOS we see a lower post-merger frequency of  2.36 kHz.
Post merger we see subdominant modes more excited, and the presence of a magnetic field excites higher modes more strongly at merger than the unmagnetised case. Immediately after merger the $(2,1)$ and $(3,3)$ modes are noticeably excited
with amplitudes $\mathcal{O}(1\%)$ the strength of the dominant $(2,2)$ mode. Notably, in the unmagnetised case, these modes are $\sim 3$ times smaller than in the magnetised configurations.
It has been proposed by \cite{Espino:2023llj} that measurements of both the $(2,1)$ and $(2,2)$ modes in the post-merger phase may provide 
a constraint on the energy density gap of a potential QCD phase transition in the merger remnant. 
This phase transition acts to weaken the one-armed spiral instability in the remnant and consequently the $(2,1)$ mode. \cite{Espino:2023llj} consider evolutions without a magnetic field, and the enhancement of the $(2,1)$ mode that we see as a result of the 
presence of a magnetic field may affect the detectability of this feature, 
though a further study with EOS incorporating phase transitions, as well as magnetic 
field evolution would be required to make a conclusive statement.
When comparing to the \LOWB~ configuration, we see that, while for the first $\sim 0.5$ms post-merger the evolution of the subdominant modes appears similar to the \NOB~ configuration, these modes are quickly excited, by $\sim 1$ms post-merger, to the level of the same modes of the fully magnetised configurations.
The \STIFF~ configuration also shows a similar excitement of these higher modes at merger to the unmagnetised SFHo case, as it is more resistant overall to the induced oscillations, due to the stiffer EOS. 

\begin{figure}[t]
  \centering
    \includegraphics[width=0.49\textwidth]{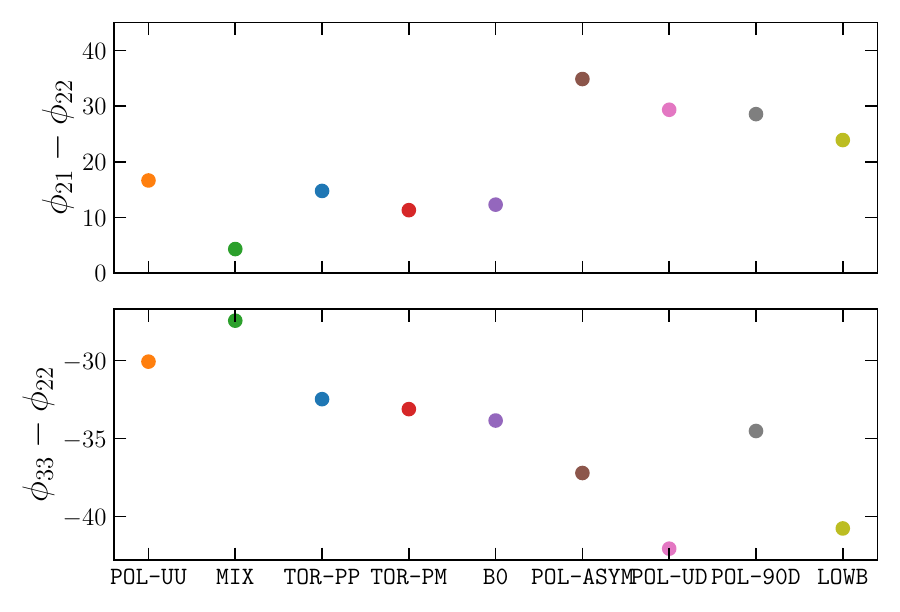}
    \caption{Upper (Lower) panel: The dephasing between the $(2,2)$ mode and the $(2,1)$ ($(3,3)$) mode $\Delta \phi_{21}$ ($\Delta \phi_{33}$). 
}
 \label{fig:gw:dephase}
\end{figure}

Post merger, the phase evolution of the subdominant GW modes for different magnetic field configurations differs. 
As discussed below in Sec. \ref{ssec:bamp}, shortly after merger we see a strong dependence of the magnetic field evolution
on the initial configuration. The differing distribution of the magnetic field, and the associated pressure and tension at this time
may have an impact on the density evolution, and consequently the GW emission in this period immediately post-merger.
By comparing the phase evolution of a given GW mode with the reference $(2,2)$ mode, we can provide a measurable impact of 
the magnetic field on the GW emission.  Specifically we  consider the $(2,1)$ and $(3,3)$ modes, since they are relatively high in amplitude 
(a factor of $\sim 50\times$ smaller than $A_{22}$), and consider their dephasing with respect to the reference $(2,2)$ mode.
At the moment of merger, denoted by the peak in GW strain amplitude, we set the GW phase to zero as a reference, and track the phase difference accumulated between the $(2,2)$ mode and these subdominant modes, up until the 
time at which the GW amplitude drops to half of its post-merger maximum. In Fig. \ref{fig:gw:dephase} we show a scatter plot of these dephasings $\Delta \phi_{\ell,m} = \phi_{\ell,m} - \phi_{22}$.

We see that, compared to the unmagnetised case \NOB, the presence of a magnetic field may 
induce a dephasing between the $(2,1)$ ($(3,3)$) and $(2,2)$ modes of up to 22.5 (8.2) radians 
by the time the GW signal has decayed to half of its post-merger peak, maximised by the \POLASYM~ (\POLUD) 
configuration, while over all different magnetic field configurations an overall difference in the dephasing 
of up to 30.5 (14.6) radians can be reached.  There is no clear hierarchy visible between the \NOB, \LOWB~ 
and remaining strongly magnetised configurations, with the \LOWB~ configuration in fact providing one of the largest dephasings with respect to the \NOB~ configuration.

We compare the dephasing effects seen in our reference SR configurations with two reference 
runs at a lower resolution for the \POLUU~ and \NOB~ configurations to estimate the error in these effects. 
For these low resolution runs
we find a difference in $\Delta \phi_{21}$ between the \POLUU~ and \NOB~
configurations of 3.0 radians, and 5.9 radians for $\Delta \phi_{33}$. The corresponding dephasings 
for our SR configurations are 4.3 and 3.8 radians respectively. These provide a relative error
estimate of $30\%$ and $55\%$ respectively for the dephasing in the $(2,1)$ and $(3,3)$ modes arising from the finite resolution.
We however see a much larger error arising from the choice of reconstruction method. In Fig. \ref{fig:gw:dephase}
we see that the dephasings of the configurations performed with the WENOZ reconstruction, and with the PPM 
reconstruction, form two distinct groups. We perform a direct comparison of the same configuration
with these two reconstruction methods in Appendix \ref{app:recon}, and quantify the error in $\Delta \phi_{21} (\Delta \phi_{33})$
as 70\% (38\%) respectively. However, the absolute size of variation in $\Delta \phi$ arising from 
the choice of reconstruction is, for the $(3,3)$ mode, larger than the range of dephasings seen above for different 
magnetic field configurations in this mode.  
The size of the variation in dephasings for different magnetic field choices for a given reconstruction choice 
is approximately constant between the two reconstruction choices considered here, however it is clear that the dephasings seen 
are highly dependent on the reconstruction routine used in the numerical evolution.

\subsection{Magnetic Field Evolution}
\label{ssec:bamp}
\subsubsection{Magnetic Field Amplification}
We now consider the global amplification of the magnetic field, and direct the reader to Paper I for a discussion of the magnetic field in the post-merger remnant and disk.
At the initial moment of merger we see the amplification of the magnetic field through the 
onset of the KHI. Here a shearing layer forms between the two stars which breaks down, forming small vortices in the fluid flow.
These vortices wind up the magnetic field lines, pinned to the fluid flow within the ideal MHD approximation, leading to an amplification of the magnetic field strength.
From high resolution simulations \citep{Kiuchi:2015sga} it is known that this instability can amplify the magnetic field energies by in excess of 6 orders of magnitude,
with the limiting factor being the resolution with which the simulations are performed. Finite resolution effects restrict the size of the vortices that can be captured, and so restrict 
the scale at which the amplification due to the KHI cannot be resolved beneath.  

\begin{figure}[t]
  \centering
    \includegraphics[width=0.49\textwidth]{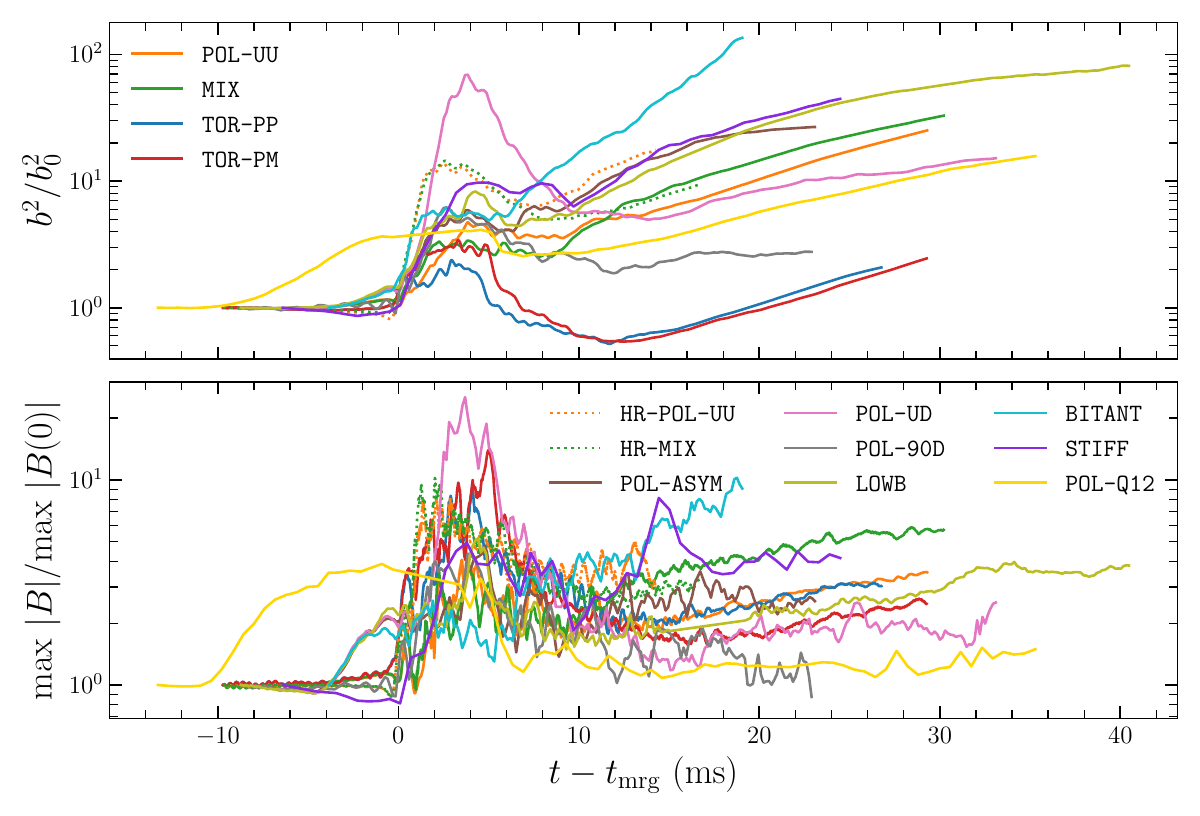}
    \caption{Upper panel: The amplification of the magnetic field energy. Lower panel: The amplification 
	of the maximum magnetic field strength. 
}
 \label{fig:bamp}
\end{figure}

We measure the dependence of this instability on our initial field geometry through first the total energy amplification 
and secondly the amplification of the pointwise maximum magnetic field strength, shown in Fig. \ref{fig:bamp}.
In the upper panel of Fig. \ref{fig:bamp} we see the amplification of the magnetic field energy as a proportion of the initial energy.
First let us focus on the amplification arising from the KHI, centered around the moment of merger $t=0$. 
We see that, at this moment, the configurations initialised with toroidal initial data see the least amplification of the magnetic field. The \TORPP~ configuration 
amplifies the least, by a factor of $\sim{}2$, while the \TORPM~ configuration amplifies slightly more, at a similar value to the \MIXED~ configuration, and 
\POLUU~ configuration, which reaches an amplification of a factor of $\sim{}5$ approximately 4 ms post-merger. We note that the higher resolution simulations 
of the \POLUU~ and \MIXED~ configurations demonstrate, as expected, a considerably larger amplification, amplifying twice as much as their lower resolution counterparts
at the moment of merger, but demonstrating qualitatively similar behaviour at the moment of merger. Since the 
\TORPP~ configuration demonstrates very little amplification, the addition of such a component to a poloidal 
component has very little effect, with the \MIXED~ configuration and \POLUU~ configuration behaving very similarly. 
Breaking the symmetry of the poloidal magnetic field configuration can however have a considerable effect. When the magnetic 
fields are totally anti-aligned, and poloidal, for configuration \POLUD~ we see the largest amplification of all,
with an amplification of the magnetic field energy by a factor of $\sim{}100$. This amplification is more even than the increased amplification 
seen by increasing the resolution of the \POLUU~ configuration. Clearly, the reversing of the orientation of the magnetic field 
in one star compared to another leads to an overall amplification of the field, with the \TORPM~ configuration also  amplifying 
considerably more than the \TORPP~ configuration, though the effect here is much less stark due to the overall lower 
amplification due to the toroidal field. 
We see in contrast that configurations \POLLR, 
\POLASYM~ undergo similar behaviour to \POLUU~ at the moment of merger.
We note also that, when normalising the overall magnetic field energy by the initial value, 
the behaviour at merger for the KHI amplification is very similar between the \POLUU~ and configuration 
\LOWB. 
As expected, the low initial value of the magnetic field, does not affect the ability of our simulation to capture the amplification due to the KHI, though of course the absolute 
value of the magnetic field does not reach the values attained for those configurations with initially much larger magnetic fields.

Immediately after the initial amplification associated to the KHI, we see an intermediate phase before the onset of a second phase of amplification, driven by winding, which lasts between $\sim$5-10 ms post-merger.
For most 
configurations the magnetic field energy remains approximately constant during this phase, with the following exceptions. All toroidally dominated runs
see a considerable dissipation of magnetic field energy between $\sim{}4-12$ms post-merger, and we see also that \POLLR~ undergoes this dissipation to a lesser extent, dropping by a factor between 2 and 4. In this sense, the orientation of the poloidal field in the orbital plane 
mirrors the toroidal configuration, also oriented in the orbital plane. Configuration \POLUD~ also sees a significant drop in magnetic field energy
after the large initial amplification, dropping down to a similar level to configuration \POLUU~ by the end of this intermediate phase.

Beginning between 9 and 14 ms post-merger, we see the onset of a winding driven amplification phase. During this phase we expect to see a linear growth in toroidal magnetic field strength,
associated to 
the overall angular momentum of the remnant winding up an overall toroidal field. We perform fits to the post-merger magnetic field strength and find that the expected linear behaviour is recovered for all configurations considered. The slope of the trend clearly 
varies between configurations, and is expected to be proportional to the angular momentum of the remnant and radial magnetic field strength. A detailed discussion of the remnant properties of these 
simulations is performed in Paper I; but here we comment that the equal mass SFHo runs fall into two major groups. A group with larger winding amplification consisting of configurations \POLUU, \MIXED, \POLASYM, \LOWB, \BIT, and a group with a lower winding amplification, consisting of configurations \TORPP, \TORPM, \POLLR~ with toroidal initial data or poloidal data aligned with the orbital plane. Configuration \POLASYM~ initially shows a large increase in energy as the winding begins, but at the end of the simulation duration, of the hydrodynamically identical full grid runs it is the \MIX~ configuration that is amplifying fastest.

We note that, with the stiffer DD2 EOS in configuration \STIFF~ the evolution of the magnetic field energy is qualitatively very similar to the 
purely poloidal configuration with the SFHo EOS, albeit with a slightly larger overall amplification, amplifying by
a factor of approximately 10 rather than 5 for the SFHo EOS. The unequal mass configuration \POLQ~ follows a similar sequence of behaviours to that of the \POLUU~ configuration, with the growth of the field beginning slightly earlier, and with a slightly reduced amplification from winding. 
As observed in Paper I, we see that, by the end of the \BIT~ configuration, it has amplified by a factor of 100 times more than the least amplified cases, the \TORPP~ and \TORPM~  configurations.
We note that in \cite{Ruiz:2020via} a direct comparison of a configuration with and without bitant symmetry is made for a different magnetic field configuration extending outside of
the star, and in contrast to our simulations, similar magnetic field amplifications are seen in both cases.

When considering the amplification of the pointwise maximum we see that, while it is still the case that configuration \POLUD~ shows the greatest amplification, the other anti-aligned 
configuration \TORPM~ shows a comparable amplification. In the KHI dominated phase, it is in fact the two toroidal configurations \TORPP~ and \TORPM~ that are the closest to configuration
\POLUD, though of course, this amplification must be highly localised in these cases, since the total energy does not show this large amplification. As in the case for the total energy, we see
a levelling off in the maximum energy for most of the configurations, while those that amplify the most drop down to comparable levels to the other configurations, in the intermediate phase before
the onset of winding. We then see an approximately linear growth of the maximum magnetic field strength in the later winding driven phase with, again, the \BIT~ configuration dominating at late times,
again a factor $\sim{} 10$ larger than the run with the least amplification 

We consider now the mechanisms that lead to such different amounts of amplification for the different magnetic field geometries. Firstly, it is clear that an anti-aligned configuration can provide 
an enhanced magnetic field amplification, as is seen by comparing configurations \POLUD~ and \POLUU, and by comparing \TORPM~ and \TORPP. 
Configurations with anti-aligned magnetic fields may be considered as candidates for enhanced reconnection of magnetic field lines due to the relative orientation of the fields.
As a proxy for this we calculate also the current  ${\bf J} = \nabla \times {\bf B}$ , and find that the hierarchy of the current matches the hierarchy of the magnetic field amplification.
In local simulations of the KHI it has been found that the growth rate of the instability is enhanced for a magnetic field that switches direction 
over the shearing interface, compared with a reference configuration with a uniform field over the interface \citep{Keppens:1999ac}. These simulations are performed 
for resistive Newtonian MHD, and here we find the same behaviour in the case of ideal GRMHD. It is suggested in \cite{Keppens:1999ac} that this enhancement arises from 
the increased reconnection, which forms field lines with increased magnetic tension, contributing to the development of the KHI. In our simulations, we also find that the hierarchy of the magnetic tension
in the star matches the hierarchy of the magnetic field and current.
Since our simulations are in ideal MHD we do not include reconnection effects, but we may expect numerical reconnection to occur due to artificial dissipation arising from our numerical 
approach. Any amplification of the magnetic field arising from this effect may be expected to compete with a loss of magnetic field energy in reconnection events, which should 
convert magnetic field energy into kinetic energy; though clearly here we see that the field amplification is increased in these cases with anti-aligned fields.

We see also that toroidally dominated configurations generate much less amplification than poloidally dominated configurations, and 
that the addition of a toroidal field to the poloidal field in configuration \MIXED, leads to relatively little change in the initial KHI amplification. In the configurations considered, the magnetic field in configurations \TORPP~ and \TORPM~ is maximised not in the equatorial plane, but at non-zero values of the $z$ coordinate. When the stars first merge, and
the shearing layer associated to the KHI first forms, it is initially formed in the equatorial plane, as this is the first point of contact between the stars. For the toroidally dominated
configurations, the magnetic field is comparatively weaker at this location, and so the amplification due to the KHI is relatively weaker. We see this feature also by considering
the strength of the magnetic field at low densities within the star. During the inspiral, we see that, at the low densities close to the surface of the star, which will make first contact between the
two stars, the magnetic field is weaker in the toroidally dominated configurations, compared to the poloidally dominated configurations. 

In the intermediate phase, we see a dissipation of magnetic field energy. This has previously been observed \citep{Cook:2023bag}, and can arise from the small scale structures formed during the KHI, 
associated to the amplified magnetic field, dropping below the grid resolution, with numerical dissipation effectively dissipating some of the amplified field.

In the winding phase we find a slightly suppressed 
growth rate for the toroidally dominated configurations, and for \POLLR, the poloidal configuration that aligns its field 
within the orbital plane. The growth rate associated to the winding is proportional to both the angular velocity of the remnant, and the strength of the radial field.
In Paper I 
we find similar rotation rates between all models in the remnant, while from Fig. \ref{fig:bamp} we see that the field strength is lesser in these configurations as the winding 
process begins. 
In comparison to the benchmark configuration \POLUU, we see that, during the winding phase the introduction of a small toroidal field in the \MIXED~ configuration appears to enhance the size of the growth, though 
this feature is clearly not robust to the increase of resolution, as we see this hierarchy reversed at higher resolution. 
The final winding driven amplification appears consistent for configuration \MIX~ at HR, while this is enhanced for \POLUU~ at HR.
In addition, the asymmetry introduced in \POLASYM~ provides an enhanced 
amplification in the winding phase. Contrastingly, the \POLUD~ configuration, which amplifies more strongly during the
KHI phase, appears to amplify less quickly during the winding phase, compared to the \POLUU~ configuration.
During the winding phase it is the \MIXED~  configuration that amplifies its pointwise maximum magnetic field the most, of all the runs with no assumed symmetry, amplifying by a factor of $\sim{}6$, compared with a factor $\sim{}4$ for the benchmark \POLUU~ configuration.

\subsubsection{Toroidal Field Evolution}
\begin{figure}[t]
  \centering
    \includegraphics[width=0.49\textwidth]{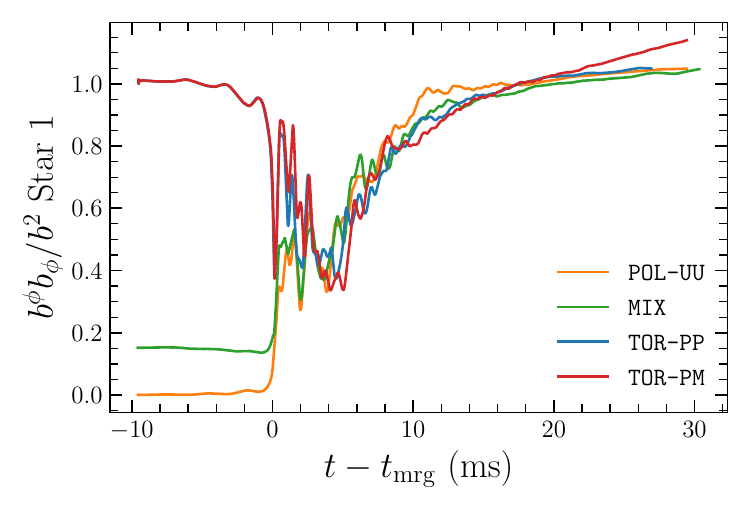}
	\caption{The toroidal magnetic energy as a proportion of the total magnetic energy of a single star. }
 \label{fig:btor}
\end{figure}

In Fig. \ref{fig:btor} we track the development of the toroidal field inside a single star during the inspiral and merger, focussing on the \MIXED~ configuration, with the \POLUU~ and toroidally 
dominated configurations for comparison.
The \MIXED~ configuration is constructed to approximate the end state of the dynamical instabilities that purely poloidal configurations such as \POLUU~ are known to suffer from, 
which generate toroidal fields. We note that Paper I contains a full discussion of the toroidal field within 
the remnant, and so here we focus on the relevance to the choice of initial data during the inspiral. 
We plot $b_\phi b^\phi$ normalised by the norm of the 4 vector $b^2$, noting that this quantity can exceed 1.
 We see that during the inspiral the toroidal component of the magnetic field within the stars is largely unchanged, and up until $\sim 2$ms before 
merger the field is the same as that set in the initial data. This suggests to us that the relatively short inspiral considered in this work, consisting of 3 orbits and lasting 10 ms is insufficiently 
long to observe the effects of the dynamically developing magnetic field instabilities expected in configuration \POLUU. In \cite{Cook:2025zzy} such instabilities are observed to saturate on the order of 25
ms, albeit for a stronger magnetic field strength, with a correspondingly faster Alfven time. For the magnetic field strengths considered in this study, one might expect these instabilities to develop 
in the pre-merger stars over the course of $\mathcal{O}$(100ms), driving the \POLUU~ configuration towards a configuration similar to the \MIXED~ configuration. At the moment of merger the poloidally 
dominated configurations begin to develop a toroidal field, while the toroidal configurations begin to lose their toroidal field, as the localised fields given by the initial configurations are 
disrupted, and replaced with the gradually forming global toroidal field, which seems to grow at an approximately universal rate over these configurations, with the field growing fastest at late times for \TORPM.

\begin{figure}[t]
  \centering
    \includegraphics[width=0.49\textwidth]{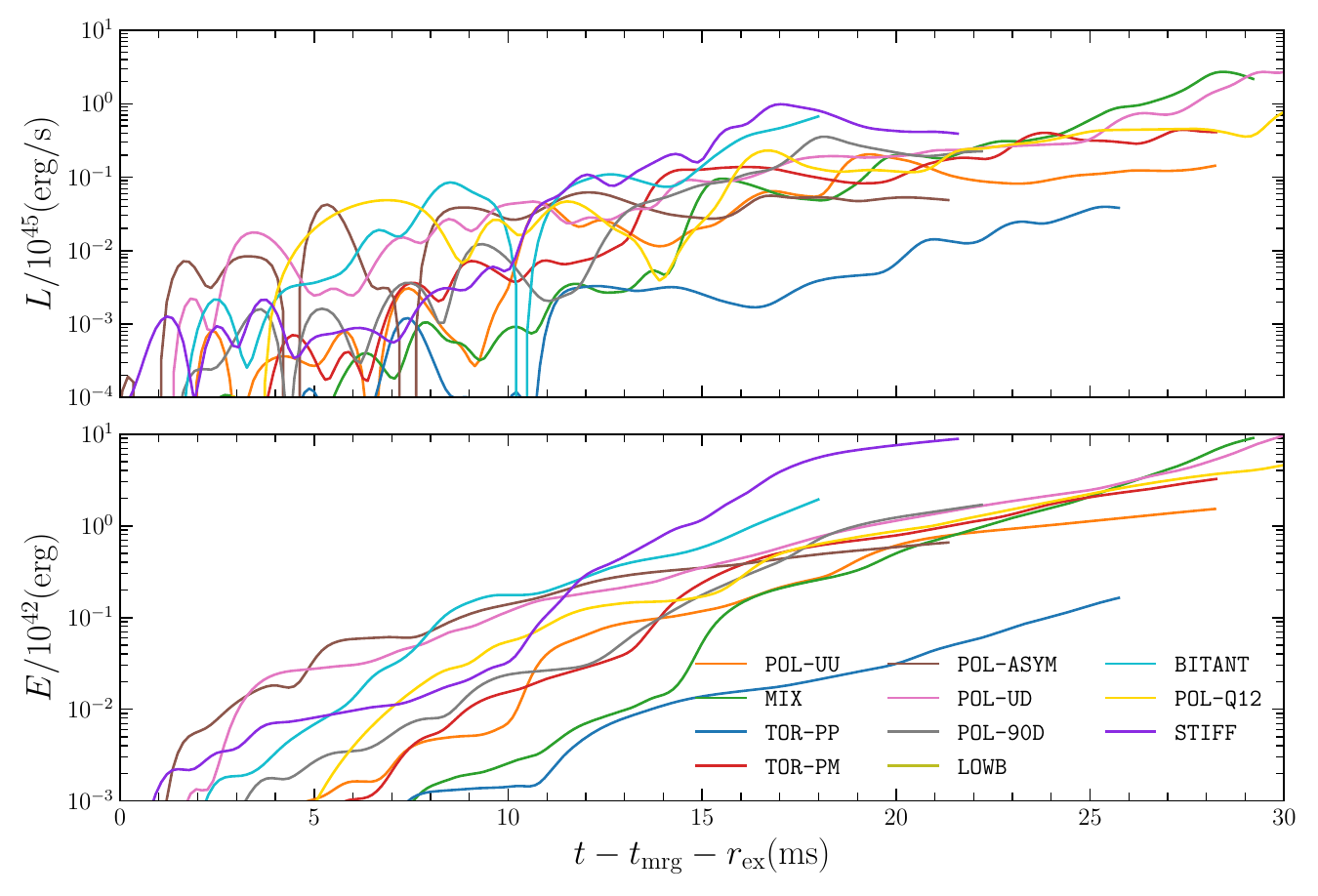}
    \caption{(Upper panel) The instantaneous luminosity of the radial Poynting flux in units of  $10^{45} \ergsec$. (Lower panel) The cumulative energy in ergs radiated through the radial Poynting flux up to time $t$. The time shown is the time in milliseconds post-merger, less the extraction radius. 
}
 \label{fig:poynting}
\end{figure}

\subsubsection{Poynting Flux}
After the merger, energy is radiated through the electromagnetic field through the Poynting flux. We calculate the luminosity associated to the radial Poynting flux, defined as,
\begin{eqnarray}
L_{\mathrm{Poynting}} = \frac{dE}{dt} &=& \int_\Omega -T^r_{t~\mathrm{EM}}r^2 d\Omega\nonumber\\ &=& \int_\Omega (b^rb_t - b^2 u^r u_t)r^2 d\Omega, \label{eq:poynting}
\end{eqnarray}
which we calculate on an extraction sphere $\Omega$ with a radius of 295~km centered at the origin. We show the luminosity of the Poynting flux, and its time integral, the radiated energy, in Fig. \ref{fig:poynting}. 

The evolution of the Poynting luminosity is broadly similar across different
magnetic field configurations. As seen in the magnetic field amplification, we see a strongly suppressed Poynting flux for configuration
\TORPP, associated to its weaker magnetic field. We note however that the other weakly amplified configurations \TORPM~ and \POLLR~ behave in a similar fashion to the other, much more strongly amplified 
configurations. By the end of the simulations the benchmark \POLUU~ configuration has the weakest Poynting flux of the configurations of similar magnetic field amplification, while the 
configuration \POLUD, with the strongest magnetic field amplification is one of several configurations with a stronger Poynting flux, though is not dominant. Instead it is the configurations that amplify
their magnetic field energies the most strongly and fastest during the winding phase that generate the strongest Poynting fluxes, the \BIT, \STIFF~ and \MIXED~ configurations. Measured 15ms post-merger these 
configurations with the strongest Poynting flux can reach luminosities a factor $\sim 10^3$ larger than that of the weakest Poynting flux given by \TORPP, while, amongst the remaining configurations
the luminosity may vary by factors $\mathcal{O}(10^2)$. In contrast, configuration \POLASYM, which has a large amplification in the winding phase, has a relatively low Poynting luminosity by the end of the simulation. We note that, at this time, its amplification due to winding has slowed, and that its peak in Poynting luminosity, $\sim 10$ ms post-merger, is coincident with its largest magnetic field amplification in the winding phase.

The \LOWB~ run has too low a Poynting flux
to be visible in Fig. \ref{fig:poynting}, though we note that, measured at the same time post-merger, its luminosity is consistently a factor $\sim{}10^{12}$ lower than that of the \POLUU~ configuration. This
factor is broadly consistent with the expected scaling of the Poynting flux with $B_0^2$, the initial strength of the magnetic field \citep{Meier:1999a,Shibata:2011fj}.

Comparing to much longer lasting, and higher resolution simulations such as those presented in \cite{Kiuchi:2023obe}, we can see that the Poynting fluxes in our simulations are consistent with these
simulations over the timespan we consider post-merger,
approaching an approximately steady phase after tens of ms. At later times, closer to 100ms post-merger, \cite{Kiuchi:2023obe} see the formation of a far stronger Poynting flux. Based on the strength of the initial magnetic field in our simulations, we may expect longer lived extensions of our simulations to reach luminosities on the order of $10^{51} {\rm erg/s}$ \citep{Shibata:2011fj}, and see an accompanying effect driving the ejection of
matter from the merger remnant.

\subsection{Symmetry}
\label{sec:sym}
High resolution NR simulations of BNS systems, evolving a range of physics, employing tabulated EOSs, 
can be incredibly expensive. In order to reduce computational cost, symmetry assumptions 
can be made, to reduce the extent of the computational domain. A natural choice for the binary 
problem is to assume bitant symmetry, assuming a symmetry over the orbital plane of the binary 
(here the $z=0$ plane), evolving only the upper half plane of the computational 
domain. The evolution is then performed using appropriate boundary conditions over the $z=0$ plane. 
Scalar quantities such as the density are set such that $\rho(z=0^+) = \rho(z=0^-)$, the velocity 
is set such that $(\tilde{u}^x, \tilde{u}^y, \tilde{u}^z)(z=0^+) = (\tilde{u}^x, \tilde{u}^y, -\tilde{u}^z)(z=0^-)$ and, 
for consistency with the standard assumption of an internal poloidal field in the star, the 
magnetic field boundary conditions are set such that $(B^x, B^y, B^z)(z=0^+) = (-B^x, -B^y, B^z)(z=0^-)$. 
We perform all of the runs considered in this paper assuming no symmetry at all, except for \BIT, 
which is initialised with the poloidal field configuration, and employs the above mentioned symmetry 
over the $z=0$ plane. We analyse the extent to which our runs with no enforced symmetry 
retain the symmetry properties of the initial data, and assess to what extent an enforced bitant 
symmetry can give a false impression of the overall magnetic field structure which develops during 
the simulation. For these purposes we define the following diagnostic quantities for the 
symmetric and antisymmetric parts of the magnetic field,
\bea
B^i_{z>0} &=& \frac{\int_{z>0} B^i \rho dV }{ \int_{z>0} \rho dV},\\
B^i_{z<0} &=& \frac{\int_{z<0} B^i \rho dV }{ \int_{z<0} \rho dV},\\
B^i_{\mathrm{sym}} &=& \frac{1}{2}\left(  B^x_{z>0}  - B^x_{z<0},  B^y_{z>0}  - B^y_{z<0},\right . \nonumber\\&& \left .B^z_{z>0}  + B^z_{z<0} \right) \\
B^i_{\mathrm{asym}} &=& \frac{1}{2}\left(  B^x_{z>0}  + B^x_{z<0},  B^y_{z>0}  + B^y_{z<0},\right .\nonumber\\&& \left . B^z_{z>0}  - B^z_{z<0} \right) \\
B_{\mathrm{sym}} &=& \sqrt{B^i_{\mathrm{sym}} B_{i~\mathrm{sym}}} \label{eq:sym}\\
B_{\mathrm{asym}} &=& \sqrt{B^i_{\mathrm{asym}} B_{i~\mathrm{asym}}}. \label{eq:asym}
\eea
In order to avoid polluting our results with 
fields supported by spurious low density ejecta, we only integrate over matter with density over 
$10^{13}\mathrm{g/cm^3}$.

\begin{figure}[t]
  \centering
    \includegraphics[width=0.49\textwidth]{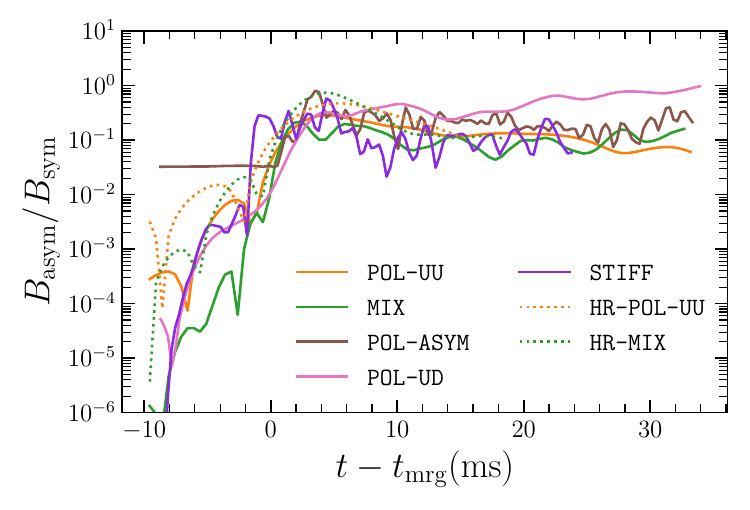}
    \caption{The ratio between the antisymmetric and symmetric parts of the magnetic field integrated over the spatial domain (Eq. \ref{eq:sym}-\ref{eq:asym}). 
}
 \label{fig:sym}
\end{figure}

In Fig. \ref{fig:sym} we present the time evolution of the symmetry of the magnetic field. 
We comment that the toroidal and \POLLR~ configurations are not adapted to the symmetry considered,
and so we do not include them here, and that
by construction, the \POLASYM~ 
configuration begins with an asymmetry. 
Clearly we see at early times the magnetic field is largely 
symmetric for the configurations adapted to the bitant symmetry, with a relative initial 
violation of symmetry on the order of one part in $10^3$ if not smaller. 
We see that asymmetry grows throughout the inspiral
phase, and that by the point of merger, the antisymmetric component of the 
field has grown to several percent the size of the symmetric field.
We see that this behaviour is largely independent of 
the specific initial configuration, with the rate of increase of antisymmetry throughout the inspiral approximately
uniform for different configurations. 
We see for the \POLASYM~ configuration, which begins at a non trivial value of the
asymmetry by construction, that, once sufficient time has passed during the inspiral for the initially symmetric configurations to 
grow an asymmetry of a similar size to that of the initial data of \POLASYM, that the \POLASYM~ asymmetry evolves in a similar manner to 
the initially symmetric runs. This suggests that the growth of the asymmetry is occurring in the same manner across all runs, but is
only visible in the \POLASYM~ configuration once it exceeds the initial asymmetry. We note that, for the runs not presented here that
are not adapted to the bitant symmetry, that while the asymmetry may be large in the inspiral due to the construction of the 
initial data, the magnitude in the post-merger phase is consistent with the other runs. This is in agreement with the 
findings of Paper I, which suggests that the toroidal field in the post-merger remnant is largely consistent between configurations.
In the immediate post-merger phase this asymmetry
grows to order $10\%$ of the symmetric field, and remains approximately at this level, similar for all configurations considered. 
At these values, approximating the true configuration with a bitant symmetry can be expected to be 
a poor model for the magnetic field structure. The most extreme configuration, \POLUD,
grows its asymmetric field up to order unity in size compared to 
the symmetric field by the end of the simulations considered, at which point a bitant symmetry 
would not be capturing the true magnetic field at all. In the benchmark \POLUU~ configuration, 
we see that the asymmetry begins to decay slightly in the post-merger phase, 
below $10\%$. This standard configuration appears to match the bitant symmetry the closest at late 
times, though the 
value of the asymmetric field is still much higher than one would hope for if a bitant symmetry run 
were expected to capture the full magnetic field dynamics. 

Despite the fact that the initial data is symmetric,
and the evolution system introduces no explicit breaking of this symmetry, we have seen a non trivial asymmetry
grow throughout our evolution, that appears to be approximately uniform in its growth rate and final
saturation value, independent of the specific initial condition. This behaviour is consistent with
the onset of a spontaneous symmetry breaking bifurcation, with the solution to the Euler Equations 
undergoing a transition to a lower symmetry than that of the initial conditions (see for instance \cite{Crawford:1991a} for a review of such phenomena). 
The growth rate of the instability
should then be driven by the growth rate of modes close to the bifurcation, independent of the initial data itself.
The imposition of bitant symmetry, such as in the \BITANT~ configuration prevents this symmetry breaking 
from occurring, and so associated phenomena will be neglected in such simulations.
In particular, in Paper I we discuss a suppressed formation of turbulent 
features in the \BITANT~ configuration. Here we see that the formation of features in the magnetic field present
in full grid symmetry, but not in the \BITANT~ symmetry can lead to a non trivial breaking of the symmetry in the magnetic field.

\subsection{Ejecta}
\label{ssec:ej}
The binary merger process ejects neutron rich matter from the post-merger remnant, which will  decay to stable isotopes, giving off a long lived electromagnetic 
signal, the kilonova \citep{Li:1998bw,Kulkarni:2005jw,Metzger:2010sy}. This process is believed to be the origin of a large proportion of the heavy 
elements found in the universe \citep{Pian:2017gtc,Kasen:2017sxr}. The presence of magnetic fields may create additional pressures in 
the post-merger remnant, which can affect the amount of material ejected from the merger, 
potentially enhancing the material undergoing r-process nucleosynthesis, and contributing to the kilonova. In 
addition, the nature of the magnetic field itself within the ejecta may influence the weak processes that are energetically favorable to 
occur during the decay of the neutron rich matter. The magnetic field configuration may also affect 
the long term evolution of the ejected matter, trapping charged particles close to the remnant, or beaming them quickly away \citep{Barnes:2016umi}.

\subsubsection{Overview}

\begin{figure*}[t]
  \centering
    \includegraphics[width=0.99\textwidth]{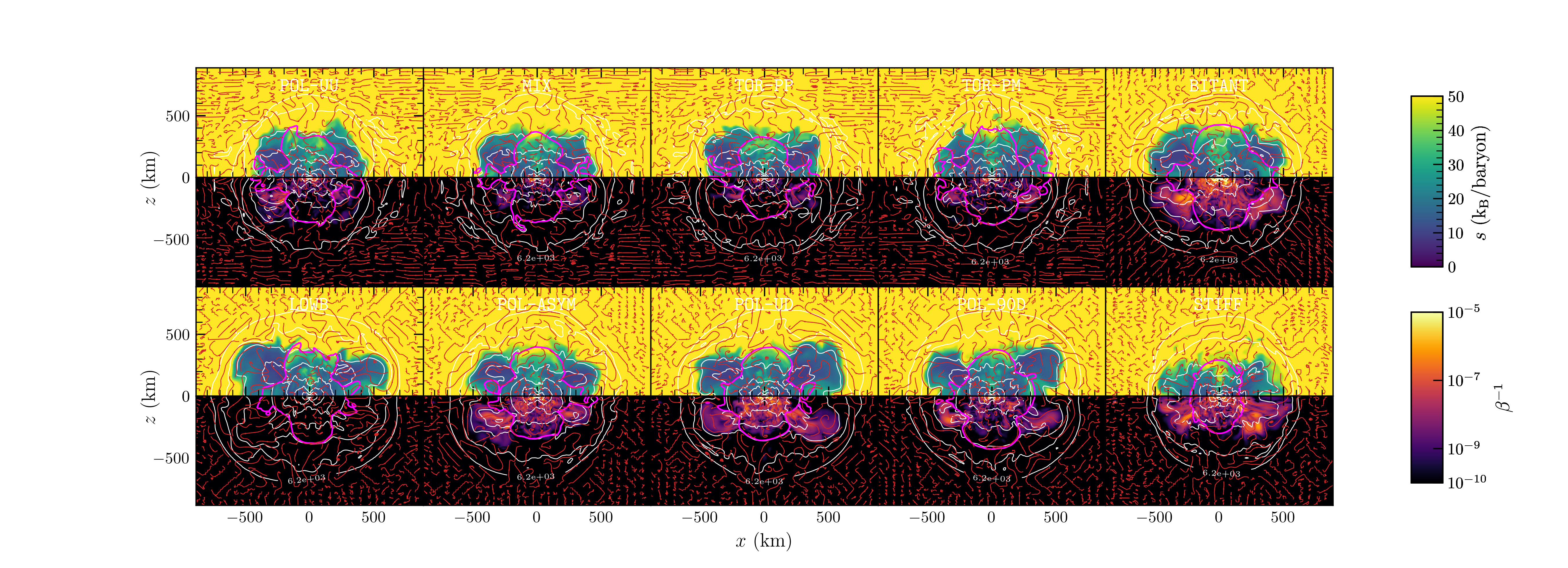}
    \includegraphics[width=0.99\textwidth]{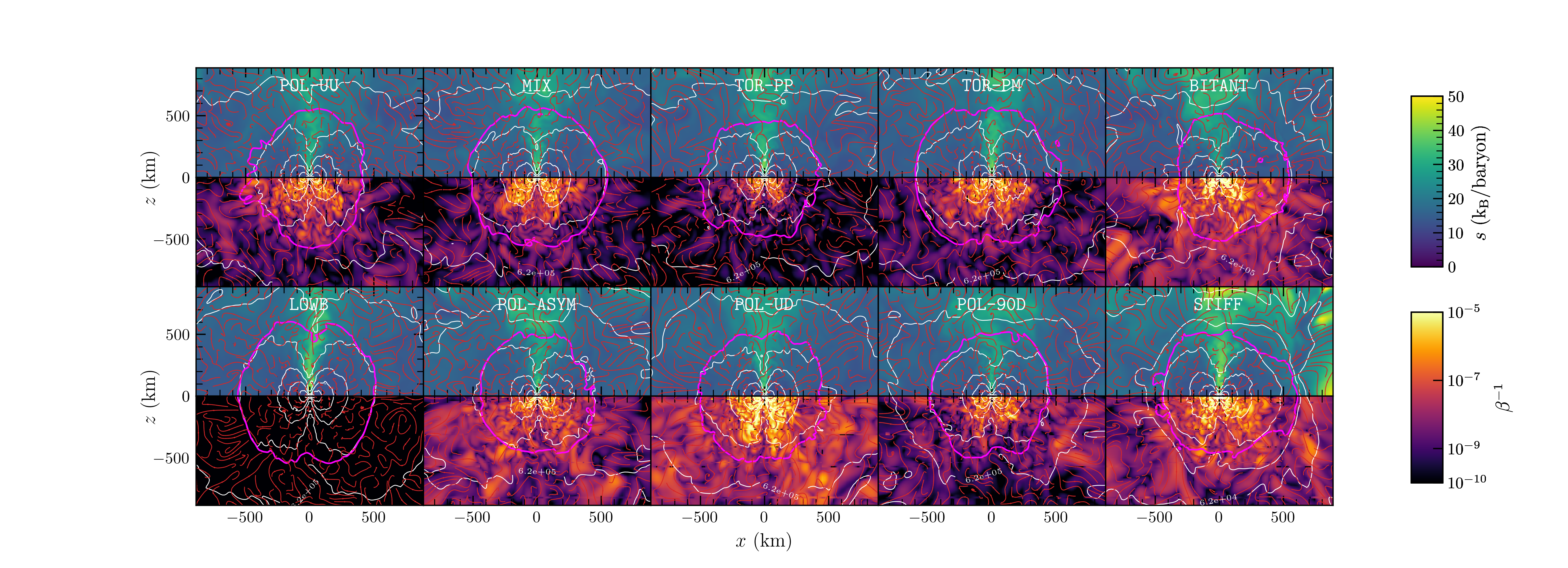}
    \caption{2D slices in $x-z$ plane demonstrating entropy per baryon in the upper half plane and $\beta^{-1}$ in the lower half plane. White contours are constant density contours, labelled in units of $g/cm^{3}$. The outermost contour is labelled, all subsequent contours increase by 1 order of magnitude moving inwards towards the remnant. The thick purple contour shows a temperature of $4\mathrm{GK}$. Magnetic field streamlines are denoted in red. Upper plot: 6.5ms post-merger. Lower plot: final time in simulation. }
 \label{fig:ej:2d_xz_a}
\end{figure*}

\begin{figure*}[t]
  \centering
    \includegraphics[width=0.99\textwidth]{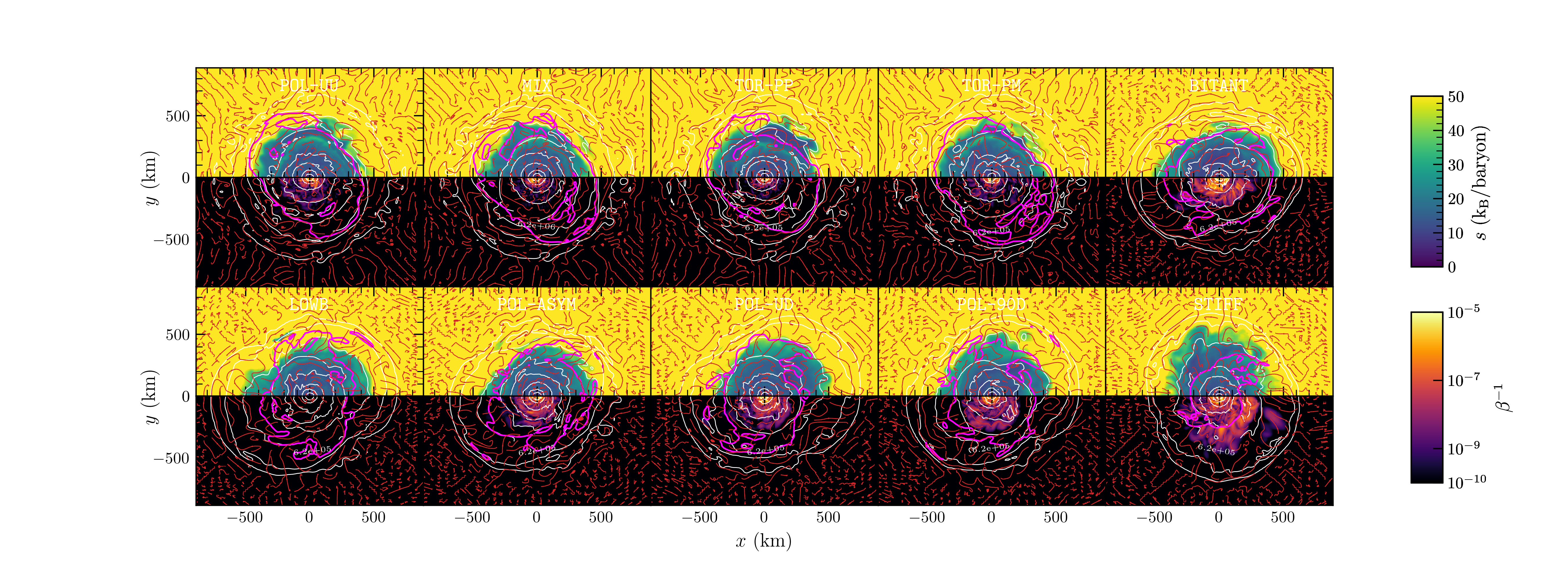}
    \includegraphics[width=0.99\textwidth]{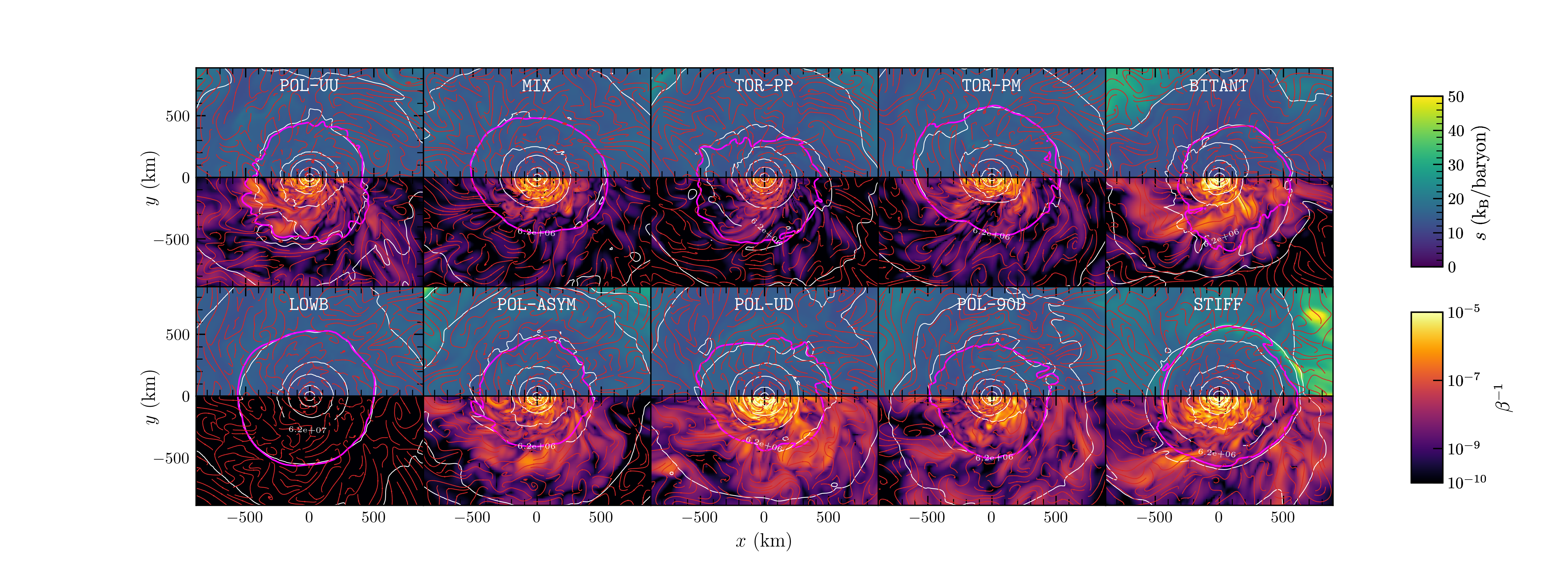}
 \caption{2D slices in $x-y$ plane. Upper plot: 6.5ms post-merger. Lower plot: Final time in simulation. Quantities plotted as in Fig. \ref{fig:ej:2d_xz_a}.}
 \label{fig:ej:2d_xy_a}
\end{figure*}

We first discuss the overall evolution of the ejected material. In Fig. \ref{fig:ej:2d_xz_a}
and Fig. \ref{fig:ej:2d_xy_a} we show the evolution of the ejected material at  large distances at
two time slices, $\sim 6.5$ms post-merger, and the final time in the simulation. We note that the final time of each run is not the same,
we refer the reader to Table 1 in Paper I for the duration of each simulation.  Due to the different reconstruction methods used (see Appendix \ref{app:recon}), the point 6.5 ms post-merger is not exactly identical between runs, however the state of the evolution is closely comparable.
In the upper panel of each plot we demonstrate the specific entropy per baryon, and in the lower panel demonstrate $\beta^{-1} = b^2/2p$, the ratio between the magnetic and fluid pressures.
In white, we demonstrate contours of density, while the thick purple contour demonstrates the temperature contour at 4 GK. This is the threshold at which ejected material has cooled sufficiently to drop out of nuclear statistical equilibrium (NSE). 

We focus initially on the evolution in the $x-z$ plane in Fig. \ref{fig:ej:2d_xz_a}. At the earlier time, we see an unphysical, high entropy, atmosphere, into which physical ejecta is 
expanding, with higher entropy material above the remnant. The 4GK temperature contour extends
outwards in the polar and equatorial regions covering material down to a density of $\sim{}6\times 
10^5 \gccm$. Between these two regions, on the diagonal of the plot, we see that there is a region of 
lower temperature and lower entropy material that has dropped below 4GK, which consistently appears 
in all configurations.
These low temperature regions correspond also to low pressure regions, and so $\beta^{-1}$ is also 
larger in these diagonal regions, with the largest $\beta^{-1}$ found for the \BITANT~ and \POLUD~ configurations. The magnetic field strength itself in this region is consistent 
with surrounding regions, and the increased $\beta^{-1}$ only arises due to the reduced $p$. 
These low pressure regions are also present in the \NOB~ configuration, though less strongly than in 
the magnetised cases. The ejecta from the  toroidal configurations  shows lower $\beta^{-1}$ 
material, due to a lower overall magnetic field strength in the ejecta, which we detail below.
From the magnetic field lines we see that the magnetic field is unstructured  in the ejected material
with no overall global configuration having formed.

By the final time we see that a funnel of higher entropy material has formed in the polar region,
with highest entropy in the \STIFF~ configuration. The 4GK temperature contour has transitioned to higher densities, approaching the $6\times10^6 \gccm$ density contour for all 
configurations with the SFHo EOS, independent of the final time of the simulation.
For the configurations \MIX~ and \POLUD, which amplify fastest and most 
strongly in the winding phase respectively, we see high $\beta^{-1}$ material emitted vertically 
outwards from the two lobes of the remnant structure in the centre, the formation of which we discuss below. The overall field structure at the end of our simulations is still not coherent
with no overall structure visible.

We see similar behaviour in the $x-y$ plane in Fig. \ref{fig:ej:2d_xy_a}, noting that higher 
$\beta^{-1}$ material is ejected further in the equatorial plane in the \STIFF~ configuration 
compared to the SFHo runs, and that at the final time higher entropy material has been ejected 
to large radii for the \STIFF~ configuration. The configurations with the largest amplification in 
the KHI phase, \POLUD~ and \BITANT, along with \POLLR, have ejected the most high $\beta^{-1}$ material to a large radius. The field itself is also unstructured in the $x-y$ plane.

It has been suggested \citep{Famiano:2020fbq,Tambe:2024usx,Kumamoto:2024jiq} that strong magnetic fields in the site of r-process
nucleosynthesis can modify $\beta$-decay rates and neutron decay rates, affecting the
relevant Urca rates, due to the quantisation of Landau levels. 
Such effects require relatively strong 
magnetic fields to be present, at least in excess of $10^{14} {\rm G}$, and for strong effects, strengths of ${\sim} 10^{16} {\rm G}$. We locate the magnetic field strength isocontour at 
$10^{14}$G in Figs. \ref{fig:ej:2d_xz_a}-\ref{fig:ej:2d_xy_a} (not pictured). This contour lies very 
close to the remnant, and is $\sim{}10$ km in radius. 
 Clearly the magnetic field strengths close 
to the remnant 
exceed these values and so, when incorporating neutrinos into such an evolution, the effects of 
magnetic fields on the
Urca rates in these regions must be incorporated. The field strengths extending outside of the 
remnant region however are insufficiently strong 
to have such an impact on these rates, as the material expands. For further detail on the field in the remnant, we refer the reader to Paper I.


\subsubsection{Impact of magnetic field configuration on ejecta}

\begin{figure*}[t]
  \centering
    \includegraphics[width=0.32\textwidth]{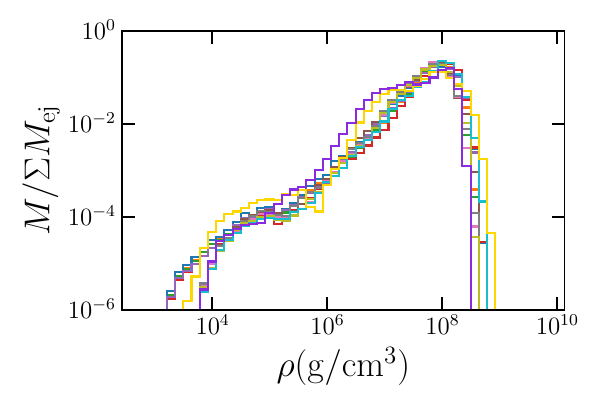}
    \includegraphics[width=0.32\textwidth]{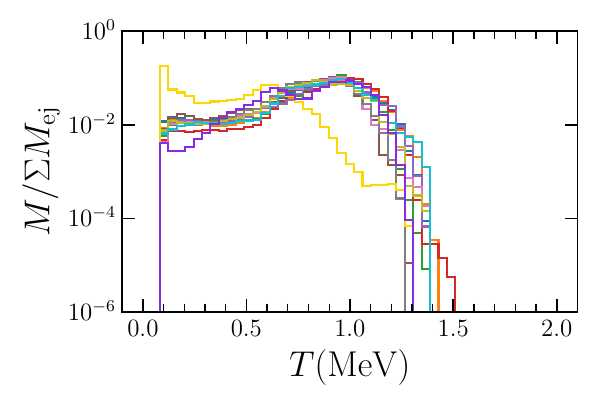}
    \includegraphics[width=0.32\textwidth]{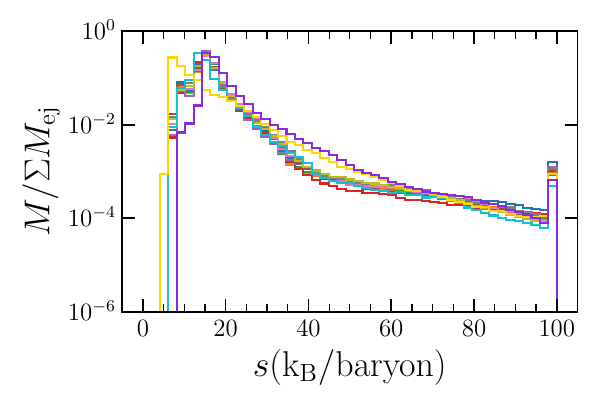}\\
    \includegraphics[width=0.32\textwidth]{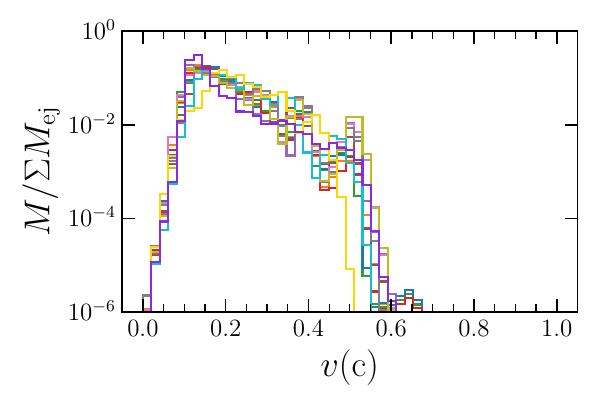}
    \includegraphics[width=0.32\textwidth]{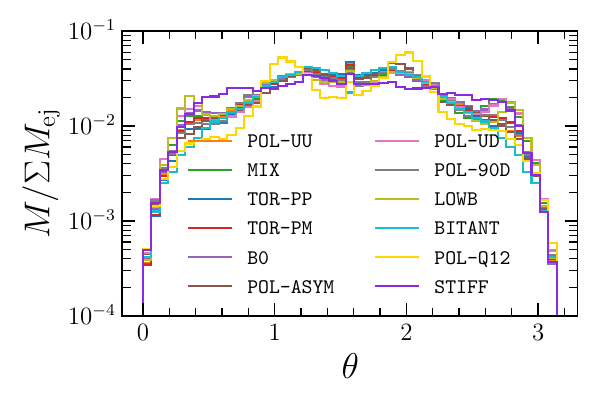}
    \includegraphics[width=0.32\textwidth]{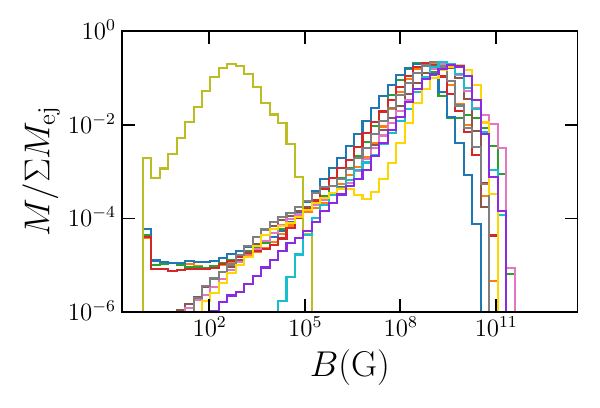}\\
    \includegraphics[width=0.32\textwidth]{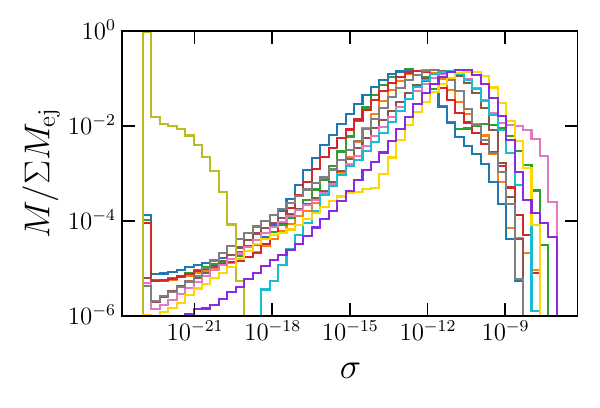}
    \includegraphics[width=0.32\textwidth]{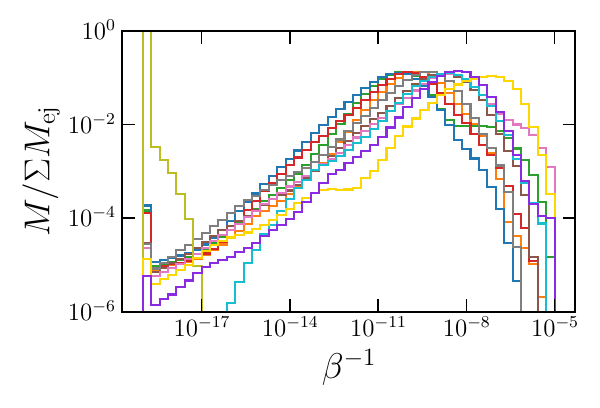}
    \includegraphics[width=0.32\textwidth]{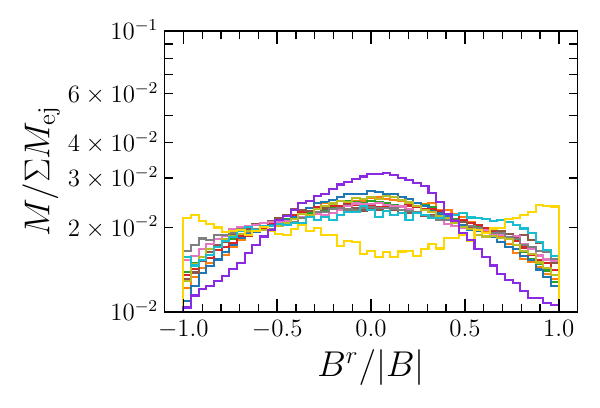}\\
    \includegraphics[width=0.32\textwidth]{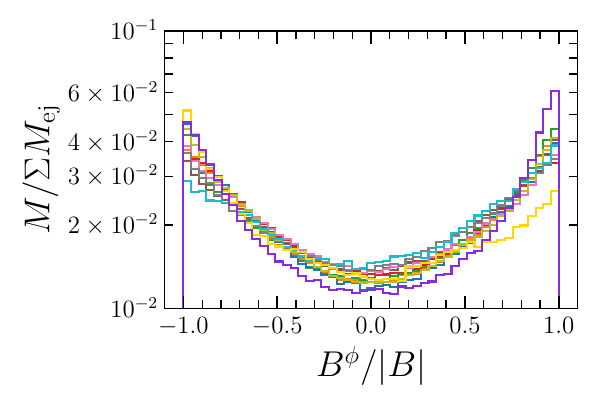}
    \includegraphics[width=0.32\textwidth]{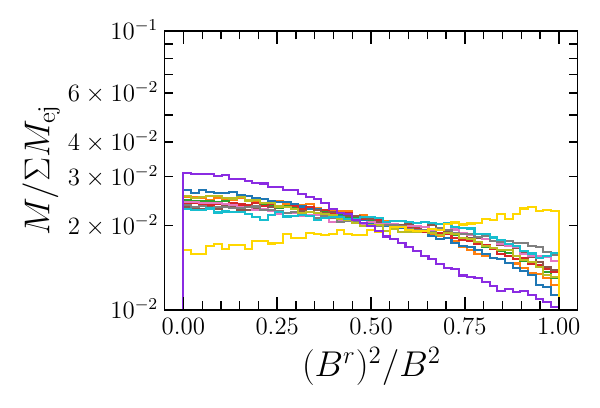}
    \includegraphics[width=0.32\textwidth]{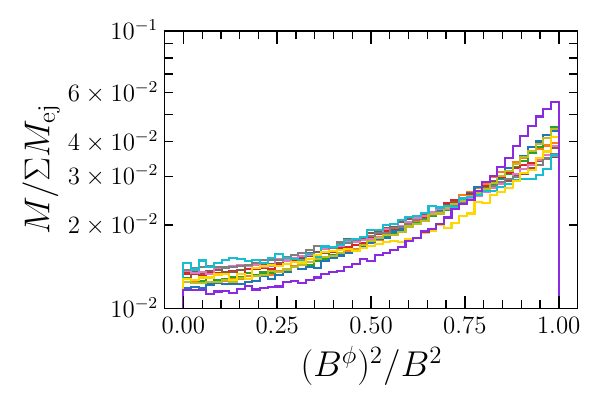}
    \caption{Histograms binned over various quantities demonstrating the cumulative proportion of the total ejected mass up to the end of the simulation, in those bins. The quantities plotted are, from left to right, then top to bottom: density, temperature, specific entropy per baryon, velocity, colatitudinal angle, total magnetic field strength, magnetisation, $\beta^{-1}$, ratio of radial magnetic field strength to total magnetic field strength, ratio of toroidal magnetic field strength to total magnetic field strength, ratio of absolute radial magnetic field strength to total magnetic field strength, ratio of absolute toroidal magnetic field strength to total magnetic field strength. Colors demonstrate different magnetic field configurations. 
}
 \label{fig:1dhisto}
\end{figure*}

\begin{figure*}[t]
  \centering
    \includegraphics[width=0.3\textwidth]{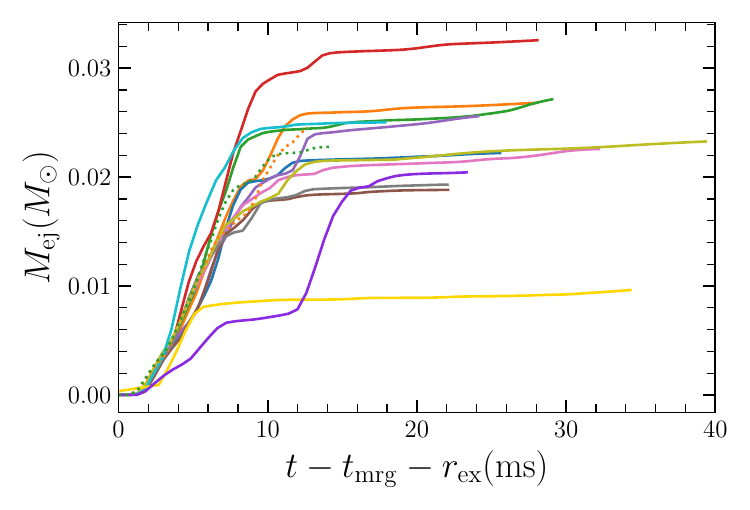}
    \includegraphics[width=0.3\textwidth]{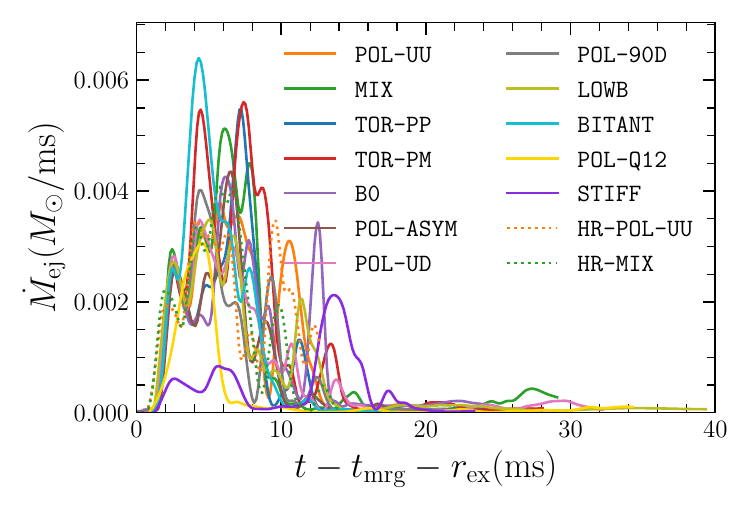}
    \includegraphics[width=0.3\textwidth]{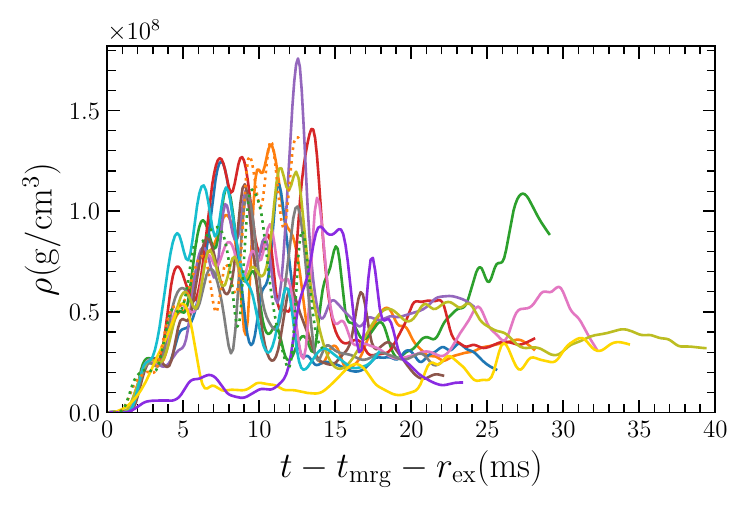}\\
    \includegraphics[width=0.3\textwidth]{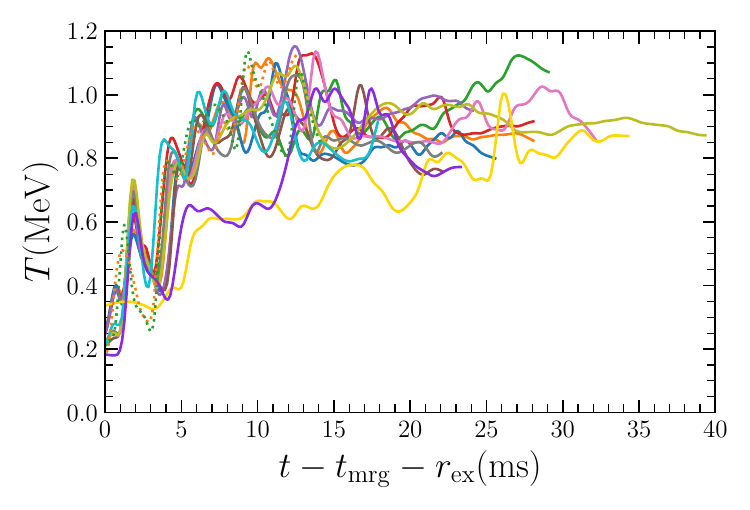}
    \includegraphics[width=0.3\textwidth]{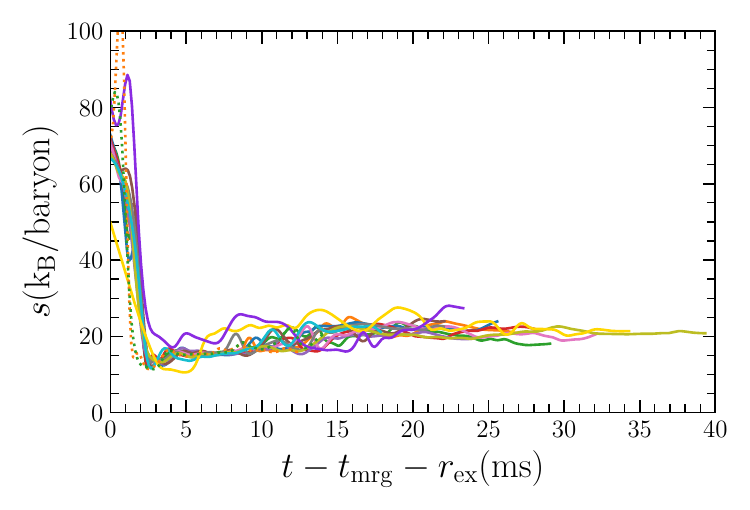}
    \includegraphics[width=0.3\textwidth]{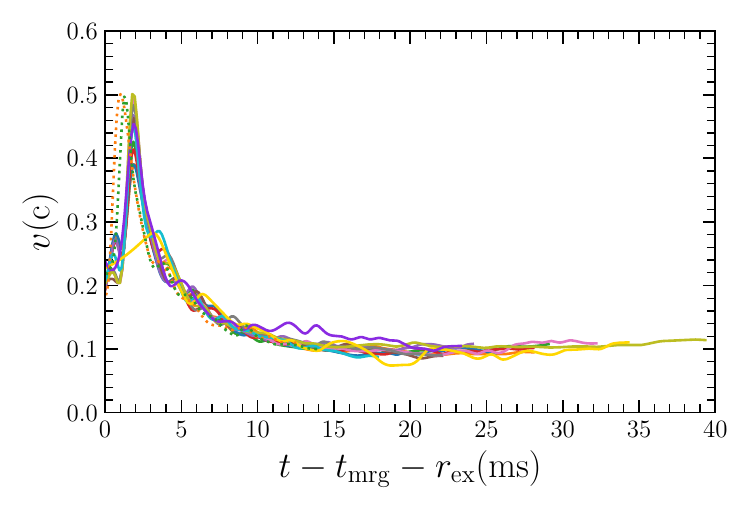}\\
    \includegraphics[width=0.3\textwidth]{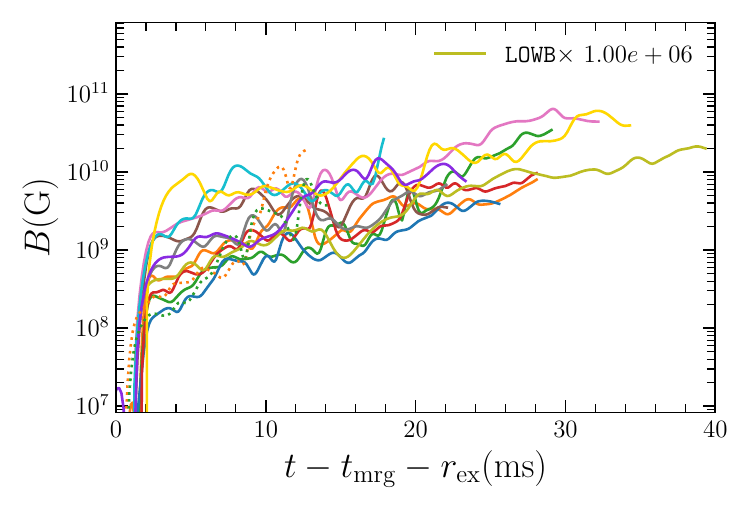}
    \includegraphics[width=0.3\textwidth]{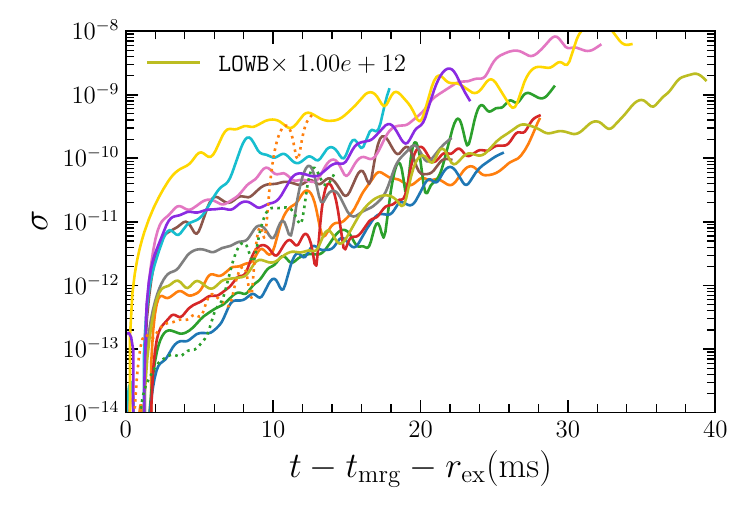}
    \includegraphics[width=0.3\textwidth]{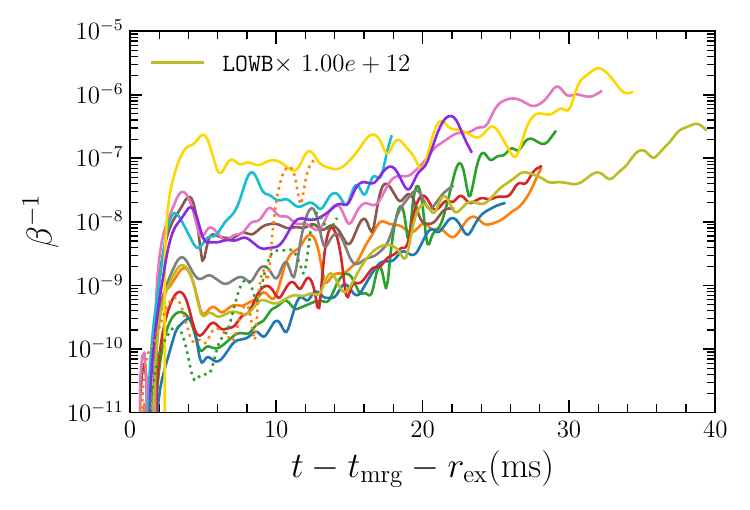}\\
    \includegraphics[width=0.3\textwidth]{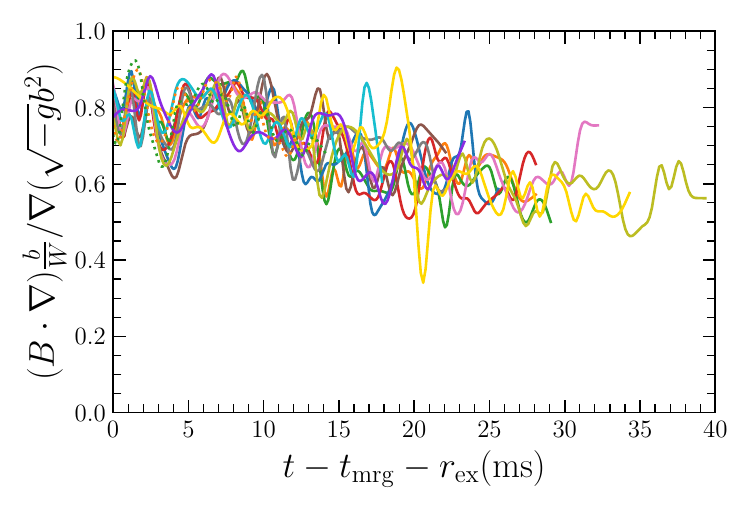}
    \includegraphics[width=0.3\textwidth]{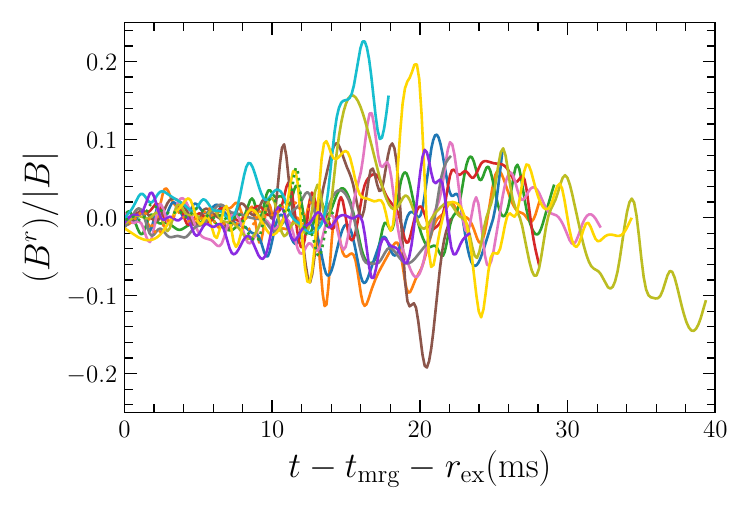}
    \includegraphics[width=0.3\textwidth]{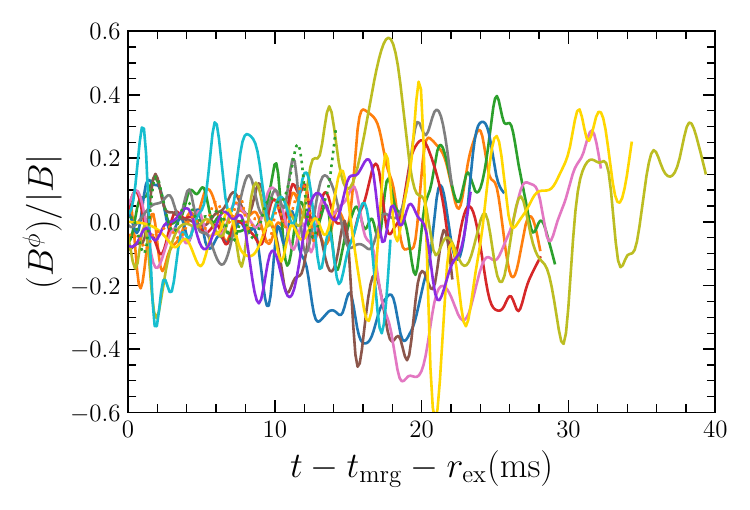}
    \caption{Upper left panel: The total ejected mass from the binary merger. Upper central panel: The ejected mass flux. Upper right panel: The average $\rho$ of the ejecta. Remaining panels show (from left to right, top to bottom) The mass weighted average of the: temperature, specific entropy per baryon, velocity, magnetic field, magnetisation, $\beta^{-1}$, ratio between magnetic tension and pressure, ratio between radial magnetic field strength and total magnetic field strength and ratio between toroidal magnetic field strength and total magnetic field strength. Quantities are plotted as a function of time after merger, less the extraction radius. Note that, when denoted in the legend, the \LOWB~ run has been rescaled for visualisation purposes. 
}
 \label{fig:aveej}
\end{figure*}

We now discuss the composition of the ejected material, and the impact of the magnetic field configuration on it. We first demonstrate the mass weighted composition of the ejecta at the end of our 
simulations in the histograms in Fig. \ref{fig:1dhisto}. To complement this discussion with a measure of the time evolution of the ejecta properties, we demonstrate also the average value of various 
diagnostic quantities within the ejecta over time in Fig. \ref{fig:aveej}. In this discussion we classify as ejecta, material that is unbound according to the Bernoulli criterion, that is material,
as extracted on a sphere of radius 295km, for which $hu_t<-1$. Here $u_t$ is the timelike covariant component of the fluid 4-velocity, while $h = 1 + \epsilon + \frac{p}{\rho}$ is the fluid enthalpy. $\epsilon$ is the fluid specific internal energy.

In the simulations performed we have neglected the effect of neutrinos, with the electron fraction remaining constant. We may expect that including these effects may affect the 
total amount of ejected material, and its thermodynamical properties, such as the entropy and temperature \citep{Radice:2016dwd}.
We nevertheless discuss such thermodynamical quantities here to flag the impact of magnetic field topology on the ejected material.

We first discuss the total amount of matter ejected from the BNS for the various magnetic 
configurations studied, demonstrated in the top left panel of Fig \ref{fig:aveej}. All equal mass 
configurations result in the range $0.019-0.033 M_\odot$ being ejected. The presence of, and particulars 
of the geometry of a magnetic field can have an effect of $\sim{}25\%$ on the overall mass ejected, with 
the unmagnetised case ejecting approximately $0.023 \Msun$. These values are consistent with the 
values seen by for instance \cite{Ciolfi:2017uak} for an equal mass configuration, albeit with a different EOS (APR4).
More recent studies such as \cite{Radice:2018pdn,Combi:2022nhg} see slightly less ejected mass, though the overall hydrodynamical evolution appears consistent,
for instance in the average entropy distribution. We note that these references use an M0 neutrino transport scheme, capturing effects neglected within our
simulations. 
For the equal mass case the extremal values are given by the \TORPM~ configuration and the \POLASYM~ configurations 
which eject the most and least mass respectively, while the \POLQ~ configuration ejects the least material overall. Note we have 
doubled the \BITANT~ mass observed in the simulation to compensate for the half grid evolution. 

In the upper central panel of Fig \ref{fig:aveej} we observe the mass flux of ejected 
material. Here we see the largest instantaneous mass flux comes in the \BITANT~ configuration, 
with larger early fluxes in the \TORPP, \TORPM~ and \MIXED~ configurations. Comparatively delayed 
ejection is seen in the \POLUU, \NOB~ and \STIFF~ configurations, which are still emitting larger 
quantities of unbound material as the  other configurations have become more quiescent.
Notably, at late times, we see continued emission from
\MIXED~ and \POLUD, more than 25~ms after merger, the configuration that is amplifying the fastest at the end of the simulation, and the configuration that has undergone the strongest amplification..

In the temperature and entropy evolutions we see the majority of configurations undergo an identical evolution, 
with entropy (temperature) enhanced (suppressed) for the \STIFF~ and \POLQ~ configurations at earlier times, between 5-15ms post-merger 
and suppressed (enhanced) at late times for the 
\MIXED~ configuration, with the ejected material averaging $\sim 17 \mathrm{k_B}$ per baryon for this configuration, compared to the majority of
configurations which average $\sim 25 \mathrm{k_B}$ per baryon, as demonstrated in the left and central plots in the second row of Fig. \ref{fig:aveej}, and corresponding histogram.

The velocity of the ejecta is also relatively consistent between configurations.
In the right panel of the second row of Fig. \ref{fig:aveej} we see that the \STIFF~ configuration demonstrates
a slightly faster average ejecta velocity compared to the other configurations, while from the histogram of velocities
demonstrated in the left panel of the second row of Fig. \ref{fig:1dhisto}, we see that the \POLQ~ configuration
contains much less fast ejecta, with almost no material ejected with velocity over 0.5$c$. We find that
the configurations with the most fast ejecta, here defined with $v>0.5c$ are the configurations  \LOWB, \NOB, and \POLUD,
with $ 2\%, 0.6\%, 0.9\%$ of the ejected mass falling into this category. 
We note that there is no clear hierarchy in the amount of ejected matter, the thermodynamics, or the velocity of the
ejecta between the \NOB, \LOWB~ and more magnetised configurations. At longer timescales, with the onset of
a magnetically driven wind, we may expect such a hierarchy to form however.

Finally, in our analysis of the hydrodynamical evolution, we discuss the angular distribution of the ejected matter. 
In all configurations the material is approximately symmetrically distributed over the colatitudinal angle $\theta$ with the variation of the mean angle between 
the configurations 0.1 radians in size, centred on $\frac{\pi}{2}$.  All configurations show a broadly consistent distribution, with the ejecta peaking in the orbital plane and just above and below, and decaying towards the polar regions. For the configurations that last the longest, there is an increase in ejected matter in the polar regions at $\sim 0.4$ and $\sim 2.8$ radians. This feature is
present also in configurations \NOB~ and \LOWB, and so we cannot ascribe this to a magnetically driven outflow. 
In configuration \STIFF, we see a much more isotropic distribution of the ejected matter than the SFHo configurations, with more matter emitted at $\theta=\sim \pi/4$ and $\sim 3\pi/4$ , where there is a dip for the softer EOS. In contrast, configuration \POLQ~ demonstrates clear peaks immediately above and below the equatorial plane, and relatively smaller ejecta mass in the equatorial plane, and in the polar regions.

\subsubsection{Magnetic field in the ejecta}

We now consider the impact of the magnetic field configuration on the magnetic field itself in the ejecta.
In the left panel, third row of Fig. \ref{fig:aveej} we demonstrate the average magnetic field strength. 
We note that the hierarchy of this strength shares some features with, but does not identically match that of the global magnification post-merger seen in Fig. \ref{fig:bamp}.
Those configurations with the strongest and fastest amplifications such as \BITANT, \POLUD, and \MIXED~ do see larger corresponding magnetic fields in the ejected material.
However, we see that the \POLUU~ configuration demonstrates a comparable magnetic field to that of the much less magnetised \TORPP~ and \TORPM~ configurations.
At ${\sim} 25$~ms post-merger, we see that the magnetic field in the ejecta may vary by over an order of magnitude, comparing
the least strong field, \TORPP~ and the strongest \POLUD, which, at its strongest, reaches $10^{11} {\rm G}$.

Other than the \STIFF~ configuration, all models have ejected the majority of the unbound material that they emit over the course of the simulation
by $\sim 15$ms post-merger (although we may expect on longer timescales more material to be ejected by magnetically driven processes, 
or neutrino driven winds were the effects included here), with the flux dropping considerably after this point.
The amplification that we see in the magnetic field however, continues approximately exponentially after this time, 
suggesting that it is low density, highly magnetised material
being emitted at this stage. This is supported by the central and right plots in the third row of Fig. \ref{fig:aveej}, in which we see 
a similar hierarchy in the magnetisation $\sigma = B^2/\rho$, and $\beta^{-1} $. We note that 
$\beta^{-1}$ still shows a fluid dominated by fluid pressure, rather than magnetic pressure.
The one major difference between these variables and the magnetic field strength can be seen for the $\STIFF$ configuration, 
which has a comparatively amplified $\sigma$ and $\beta^{-1}$ at early times, which arises from the comparatively low density and
low temperature in the material ejected over the first 10ms post-merger in this configuration, seen in Fig. \ref{fig:1dhisto}.

In Fig. \ref{fig:1dhisto} (second row, right panel) we see an
asymmetric distribution of the magnetic field in the ejected matter peaked towards the higher values of the magnetic field, with the peak varying by 2 orders of magnitude, between $10^8$ and $10^{10}$G 
(with similar behaviour for $\sigma, \beta^{-1}$), with more than $10\%$ of the ejected matter at this peak value. At values of the magnetic field below the peak we see broadly similar behaviour
across configurations, with the exception that the mass at these values is enhanced for the two toroidal configurations \TORPP~ and \TORPM, and suppressed for the \STIFF~ and \POLQ~ configurations, the distributions of which are notably shifted to higher values of the magnetic field. At higher values of the magnetic field, we see that \POLQ~ shows a narrower distribution of magnetic field strengths than the equal mass configurations.  
The shape of the distribution at magnetic field values above the modal value are however impacted by the initial configuration. 
As seen in the average values in Fig. \ref{fig:aveej} (third row, left panel), there is more ejecta at higher values of the magnetic field for the \POLUD, \MIXED, \STIFF, \POLQ~ and \BIT~ configurations. As the ejecta continues to evolve, the \POLUD~ configuration begins to eject more and more highly magnetised material, with the histogram biasing further to higher values over time. In the \MIXED~ configuration however, we see that more material is ejected slightly below the maximum magnetic field strength, with a flat feature developing in the magnetic field strength histogram at approximately $10^{10}$G.  This flat feature can also be seen more clearly in the magnetisation at $\sigma \sim 10^{-10}$ (and in the \POLUD~ configuration at $10^{-9}$), as well as in $\beta^{-1}$ at $10^{-7}$ ($10^{-6}$). 

The impact of the magnetic field on the transport of linear 
momentum can be separated into two contributions, that of magnetic pressure, proportional to the gradient of the comoving 
magnetic field strength $\nabla (\sqrt{-g} b^2)$, which acts to separate tightly spaced magnetic field lines; 
and that of magnetic tension, proportional to the directional derivative of the magnetic field along the magnetic field lines $(B\cdot\nabla)\frac{b}{W}$, which 
acts to straighten curved field lines. In the lower left panel of Fig. \ref{fig:aveej} we show the ratio between these quantities to estimate the more powerful effect in the ejection 
of unbound material. Here we see that, at early times, the quantities have comparable contributions in the ejecta, but that the contribution of
magnetic tension gradually decays post-merger, leaving magnetic pressure the dominant contribution to the ejecting force by a factor of
$\sim 2$, 30ms post-merger. This behaviour seems largely universal, and independent of the initial magnetic field configuration.

\begin{figure*}[t]
  \centering
    \includegraphics[width=\textwidth]{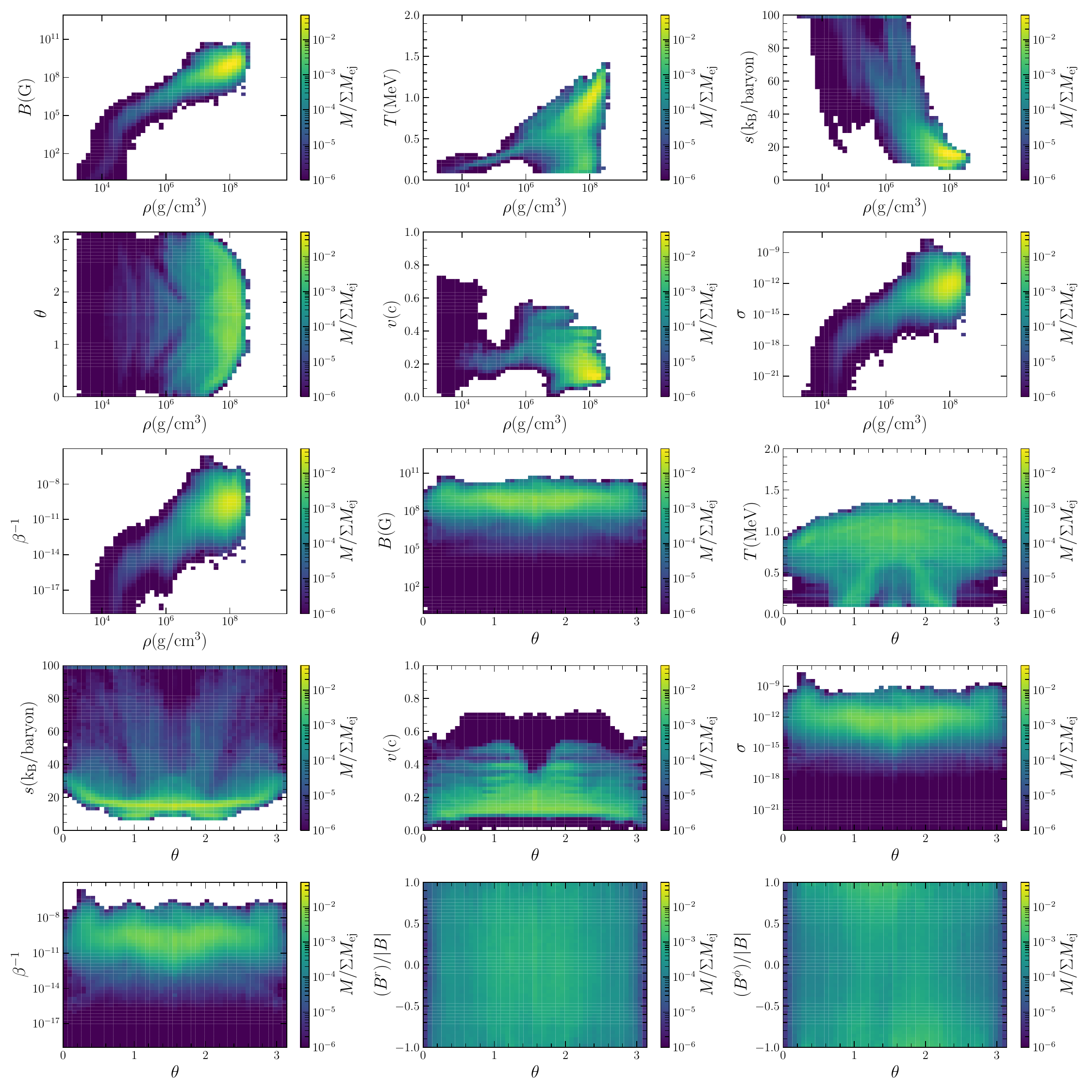}
     \caption{2D histograms demonstrating the make up of the cumulative ejected matter at the end of the \POLUU~ simulation. From left to right, top to bottom, the histograms demonstrate the following pairs of variables on the $(x,y)$ axes: $(\rho,B),(\rho,T),(\rho,s),(\rho,\theta),(\rho,v),(\rho,\sigma),(\rho,\beta^{-1}),(\theta,B),(\theta,T),(\theta,s),(\theta,v),(\theta,\sigma),(\theta,\beta^{-1}),(\theta,B^r/|B|),(\theta,B^\phi/|B|)$. The color shading demonstrates the cumulative mass ejected normalised by the total mass ejected. 
} 
  \label{fig:2dhisto_all_pol}
\end{figure*}

We now discuss the distribution of the magnetic field as a function of the ejecta density, and the angle of emission.
As a reference, we demonstrate the full hydrodynamical and magnetic state of the ejected material for the \POLUU~ configuration as a function of the
density and the angle $\theta$ in Fig. \ref{fig:2dhisto_all_pol}.
As the ejecta evolves, the ejected material transitions to larger magnetic fields, and correspondingly higher magnetisations and values of $\beta^{-1}$
, in line with the increase in density. 
At later times 
we see a steady increase in the matter ejected at higher magnetic field strengths ($\sim{}10^{11} {\rm G}$) 
at densities below the maximum, at $\sim{}2\times 10^7 {\rm g/cm}^3$. This leads to the small bump at this  
density that can be seen in the histograms for these magnetic field quantities, especially $\sigma$ and $\beta^{-1}$. 
This corresponds to 
more strongly magnetised matter outflowing from the increasingly magnetised merger remnant, flowing
into lower density material. Observing the angular distribution of these magnetic field dependent 
quantities, we see that the majority of the matter, distributed close to the equatorial plane, has 
approximately constant magnetic field strength, $\sim 10^9 G$, while the increase in magnetic field 
strength in slightly lower density material observed above occurs in the polar regions, corresponding
to the small bump in all magnetic field quantities seen in the polar region above the orbital plane.

\begin{figure*}[t]
  \centering
    \includegraphics[width=0.99\textwidth]{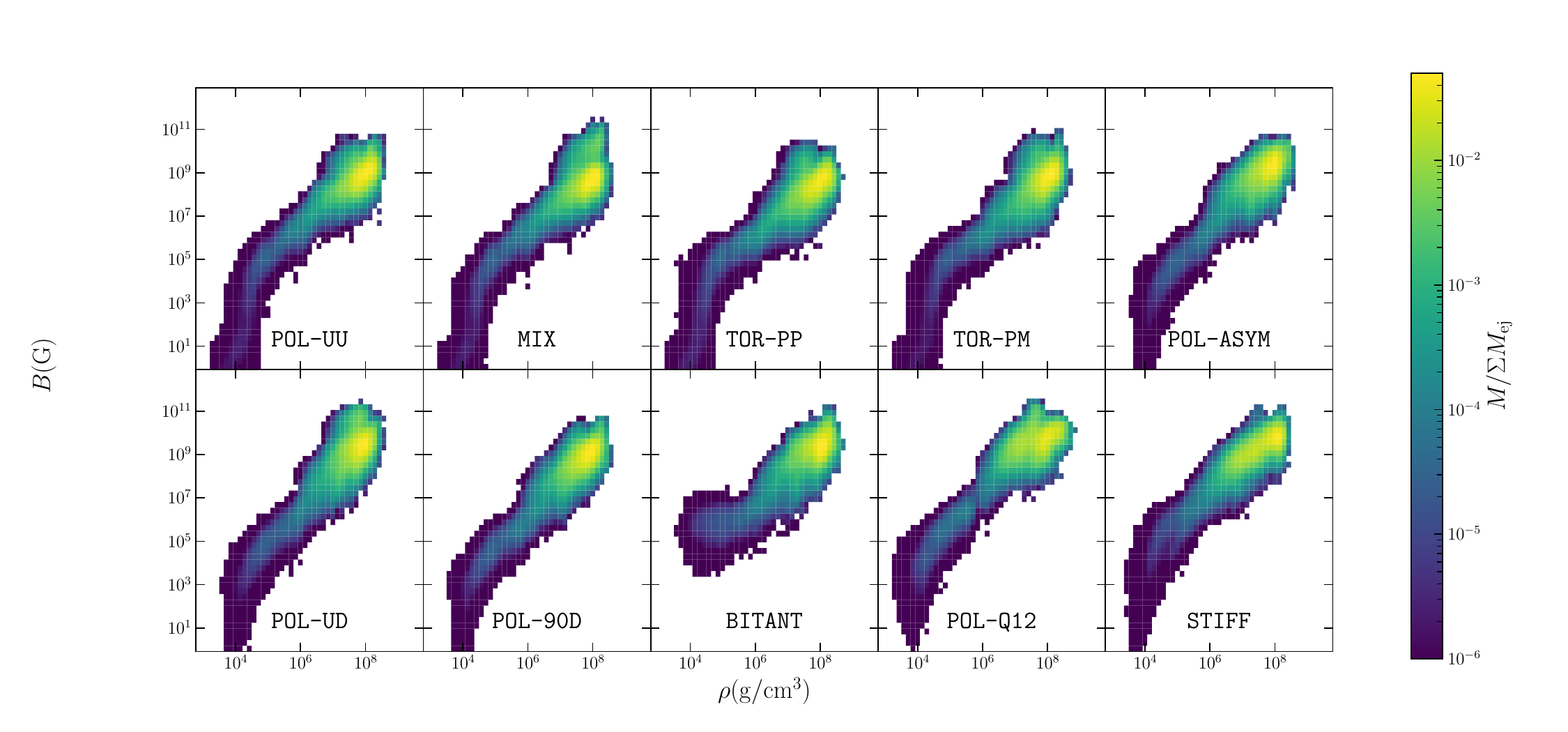}
     \caption{2D histograms of magnetic field strength against density for magnetised configurations. The color shading demonstrates the cumulative mass ejected normalised by the total mass ejected for each run. 
} 
  \label{fig:2dhisto_rho_b}
\end{figure*}

\begin{figure*}[t]
  \centering
    \includegraphics[width=0.99\textwidth]{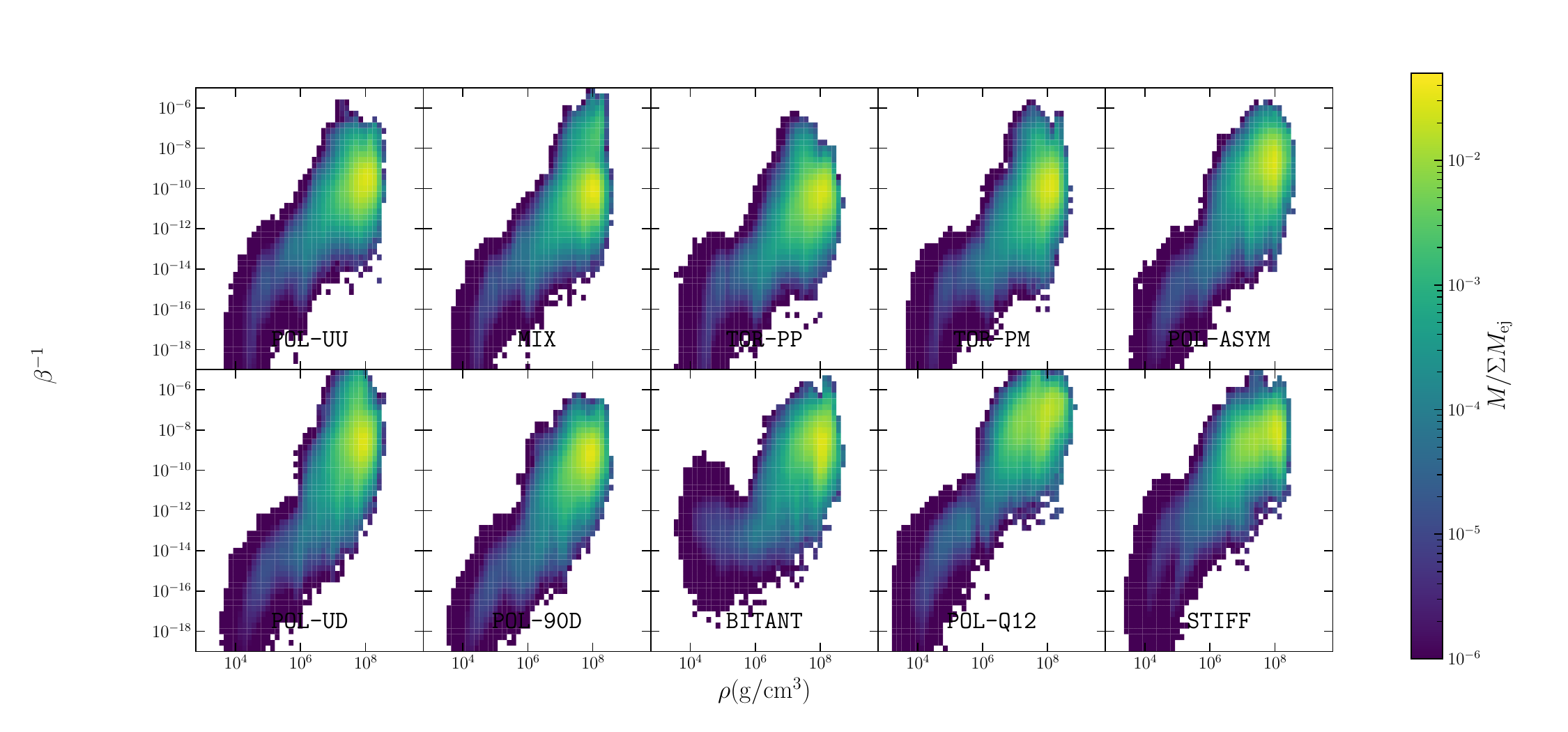}\\
     \caption{2D histograms of $\beta^{-1}$ against density for magnetised configurations. The color shading demonstrates the cumulative mass ejected normalised by the total mass ejected for each run. }
  \label{fig:2dhisto_rho_beta}
\end{figure*}

In Fig.~\ref{fig:2dhisto_rho_b} we see the distribution of the magnetic field against the density for
all configurations. 
The overall shape of these histograms is consistent over the different configurations. We see a 
mild difference in the lower density tails of pre-merger unphysical material which we ascribe to 
differences in reconstruction methods (see Appendix \ref{app:recon}), but overall see the same evolution 
from low density, less magnetised material, to higher density, more magnetised material in all 
configurations. 
In the configurations which show the 
greatest magnetic field enhancement in the ejecta, namely the \MIX, \POLUD~ and \POLQ~ configurations
we see the bump feature previously seen in the \POLUU~ ejecta clearly enhanced. Here we see for 
these configurations, at densities below the maximum in the ejecta, a subdominant 
(in terms of mass ejected) component at higher values of the magnetic field strength. This feature 
appears to be the source of the enhanced magnetic field strength in the \MIX~ configuration, since
the majority of the ejecta for \MIX~ is centred around a lower average magnetic field strength than 
\POLUU, however, the extent of this bump reaches higher magnetic field values than are seen in \POLUU.
In contrast, the majority of the ejecta in the \POLUD~ and \POLQ~ configurations can be found at comparatively higher values of the magnetic field,
and so the increased magnetic field is a global feature, rather than localised to this bump. 
We also note that the high magnetic field 
strength in the \BITANT~ configuration does not come from the development of such a 
feature, and simply comes from the overall magnitude of the magnetic field in the ejecta being larger.
In Fig. \ref{fig:2dhisto_rho_beta} we see the distribution of $\beta^{-1}$, which demonstrates a
similar hierarchy to that of the magnetic field strength, with a similar bump feature developing below the maximum $\rho$ at high $\beta^{-1}$. 

\begin{figure*}[t]
  \centering
    \includegraphics[width=0.99\textwidth]{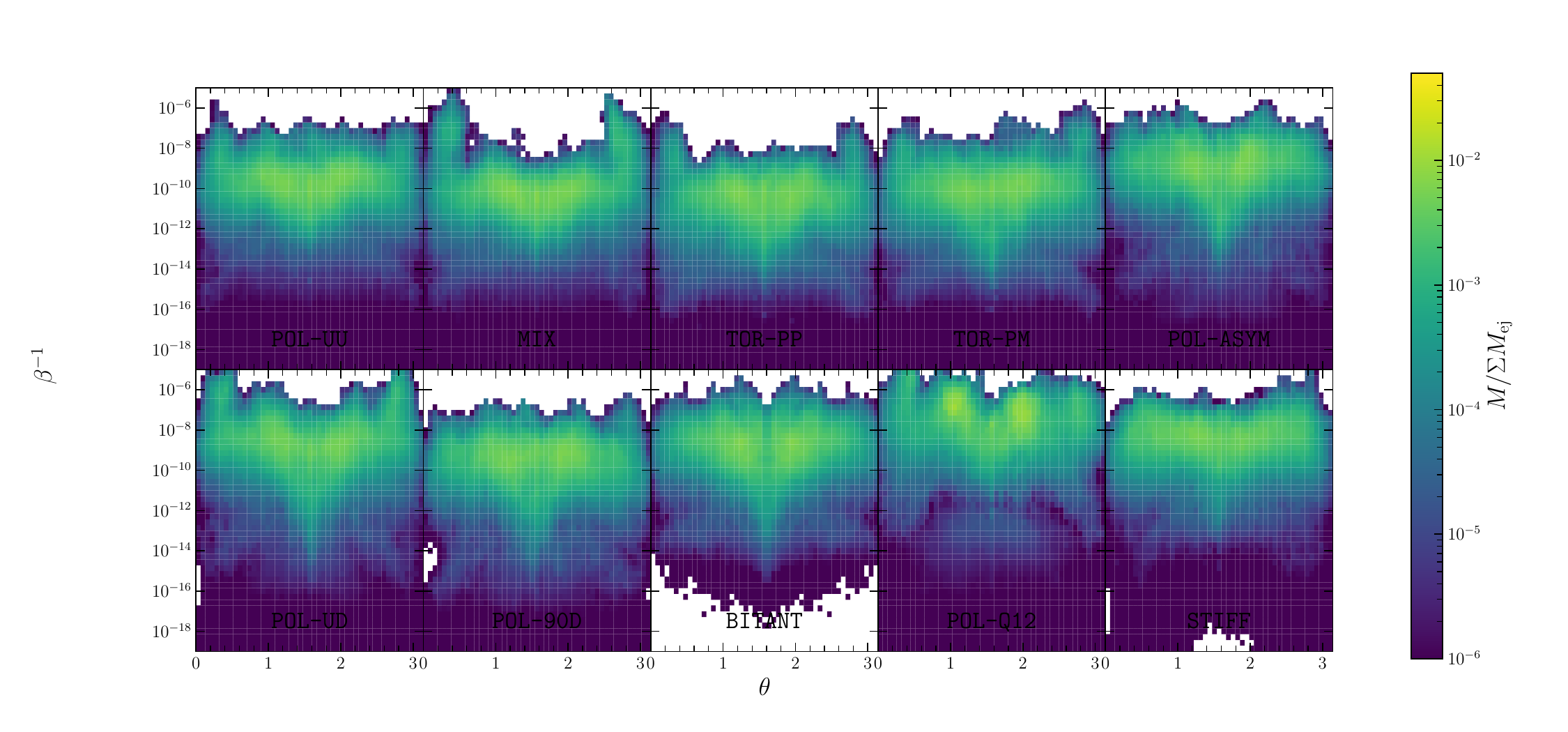}
     \caption{2D histograms of $\beta^{-1}$ against $\theta$ for magnetised configurations. The color shading demonstrates the cumulative mass ejected normalised by the total mass ejected for each run. }
  \label{fig:2dhisto_theta_beta}
\end{figure*}

In Fig. \ref{fig:2dhisto_theta_beta} we demonstrate the angular distribution of the 
magnetic field, using $\beta^{-1}$  as a proxy for this (though we note that $B$ shows qualitatively 
the same behaviour). Here we find that the configurations that have the strongest magnetic fields in the ejecta, 
forming the bump feature discussed above, correspond to those forming 
a strong field in the polar regions, as can be seen in the \MIX, \POLUD~ and \POLQ~ configurations.
We see also that the two toroidally dominated configurations are forming clear features in the polar regions, at higher values of $\beta^{-1}$. These features are also present for the toroidal configurations in the magnetic field strength, though at notably lower absolute values of the field strength than the \MIX~ configuration.

\begin{figure}[t]
  \centering
    \includegraphics[width=0.48\textwidth]{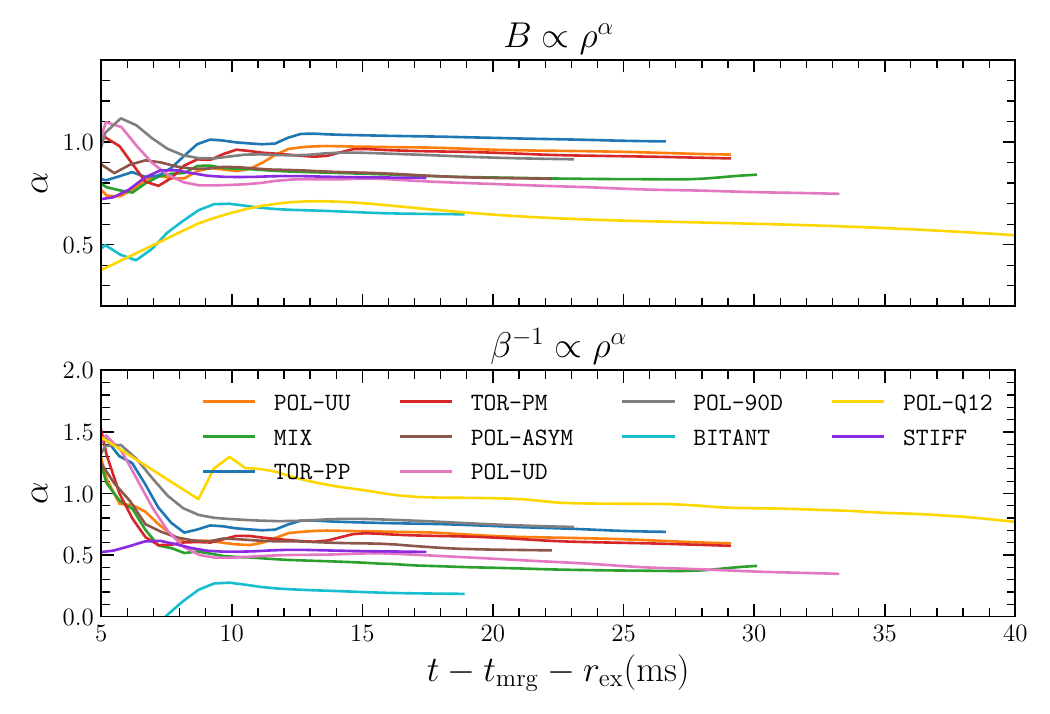}
    \caption{Upper (Lower) panel: Evolution of scaling exponent $\alpha_i$ in ejected material for power law fit $B = A_1 \rho^{\alpha_1}$ ($\beta^{-1} = A_2 \rho^{\alpha_2}$). The evolution of the exponent $\alpha$ is shown as a function of time after merger, less the extraction radius. } 
 \label{fig:powerlawexp}
\end{figure}

In Figs. \ref{fig:2dhisto_rho_b}-\ref{fig:2dhisto_rho_beta} we have seen the relationship between density and magnetic field 
strength and $\beta^{-1}$. Neglecting the initial low mass, unphysical, ejecta that arrives at the extraction radius pre-merger, 
 the ejecta evolution follows the approximately straight line in 
these log-scaled histograms visible from the bottom left to top right of these plots. 
Once the initial ejected material has passed through the extraction radius, we see the material 
slowly begin to move to larger values of the magnetic field as time evolves, as discussed above. 
To model this evolution we fit a power law 
relationship of the form $B = A_1 \rho^{\alpha_1}$ and $\beta^{-1} = A_2 \rho^{\alpha_2}$, obtaining a least squares fit for the parameters $A_i$ and $\alpha_i$ to the mass weighted distribution of ejected material.
We perform such fits for all the considered magnetic field configurations, and show the 
time evolution of the exponent $\alpha_i$ in Fig. \ref{fig:powerlawexp}. Here we 
see that the magnetic field configuration in the initial star has a non-trivial effect on the
magnetic field distribution as a function of the density in the ejecta. Up until $\sim{}$5ms post-merger our fits are affected by the unphysical pre-merger ejecta, but after this point we see that the
scaling exponent becomes approximately constant, with a mild decreasing trend, as the magnetic field
strength at lower densities gradually increases, as more strongly magnetised material is ejected. 
We can see that the exponent ranges from $\sim$ 1, demonstrating a linear relationship, for the 
configuration that amplifies its magnetic field the least, \TORPP; down to $\sim0.75$ for the full grid configuration
that has most strongly amplified, \POLUD, by the end of the 
considered simulation length; and reduces down to 0.55 for configuration \POLQ. 
Again, we find that the \BITANT~ configuration, with its enhanced 
amplification, shows a much weaker dependence, with an exponent of $\sim 0.6$ seen at the end of the simulation.
Generally, the configurations that have amplified the most show the lowest values of $\alpha$. 
The \LOWB~ configuration scales similarly to \POLUU, with the overall coefficient $A_i$ smaller.
 An overall similar hierarchy is seen for $\beta^{-1}$, though with the 
exponents smaller, demonstrated in the lower panel of Fig. \ref{fig:powerlawexp}.
Here the configuration \POLQ~ shows a clearly different behaviour to the equal mass configurations,
with one of the lowest exponents for the scaling of the magnetic field strength, but the highest for $\beta^{-1}$. This
feature arises due to the effect of the initial low density ejecta present for the unequal mass configuration, 
corresponding to the low density excess seen in the density histogram for \POLQ~ in Fig. \ref{fig:1dhisto}. Equivalently,
this can be seen in the comparatively high value of $\beta^{-1}$ in the low density tail  ($\rho \lesssim 10^6g/cm^3$) of
the 2D histogram for \POLQ~ in Fig. \ref{fig:2dhisto_rho_beta}.

\subsubsection{Magnetic field structure}

We now discuss the magnetic field orientation in the ejected material, by demonstrating the proportion of 
the magnetic field strength in the radial and toroidal components, respectively $B^r/|B|$ and $B^\phi/|B|$. 
The average of the radial and toroidal components oscillate around zero, with the magnitude of the oscillations, 
especially for the radial component strongly increasing after 15ms post-merger, as seen in the central and right 
panels of the bottom row of Fig. \ref{fig:aveej}. The size of the oscillations of the toroidal component are notably 
always larger than that of the radial component, reaching, at their largest $60\%$ of the total magnetic 
field strength, while the radial component only ever reaches at maximum $20\%$, and is mostly below $10\%$. 
By the end of the simulation duration considered there is no clear net radial field, nor a global toroidal field present in the ejected material. 
We see this feature reproduced in the histograms for these quantities, in the right panel of the third row and left panel of the bottom row of
Fig. \ref{fig:1dhisto}, with the relative strength of the magnetic field in each component shown in the final two panels, 
with the distributions approximately flat. The majority of the models, with 
the SFHo EOS and equal mass, show a slightly more toroidally dominated field, with the radial field distribution peaked at 
0.
The toroidal field contribution is peaked at $\pm 1$, with the distribution approximately symmetric, with the mean between $\pm 0.05$ for all but the \POLQ~
 configuration, with mean at -0.09. These features are enhanced for the \STIFF~ configuration which is more 
strongly peaked at 0 radial component, and a co-aligned (with the angular momentum)  toroidal component ($B^\phi/|B| = 1$). The \POLQ~ 
configuration demonstrates a much flatter distribution of the radial component, with the toroidal 
component mildly favouring a counter-aligned field.
In contrast, the \BITANT~ configuration more strongly enhances the
radial magnetic field and suppresses the toroidal field than the full grid configurations, 
though this effect is relatively small and
all field components appear to be approximately equal in significance.

The orientation of the magnetic field in the ejected material may play an important role in its evolution, and have a non trivial impact on the subsequent electromagnetic signature of the
kilonova. In \cite{Barnes:2016umi} the magnetic field in the ejected material is modelled with 3 representative orientations, a purely radial, purely toroidal, and ``random'' configuration; finding that a toroidal 
field may ``trap'' charged particles, giving them more time to thermalise, whereas a radial field may beam these 
particles outwards, reducing the thermalisation  of the ejected material. Different choices of such models of the 
magnetic field may alter the thermalisation efficiency of the ejecta significantly, and consequently impact luminosity 
predictions for the kilonova light curve. The magnetic field configurations observed here suggest that there is no 
global radial or toroidal field found in the ejecta and that the ``random'' model of \cite{Barnes:2016umi} is a 
reasonable choice.

\begin{figure}[t]
  \centering
    \includegraphics[width=0.48\textwidth]{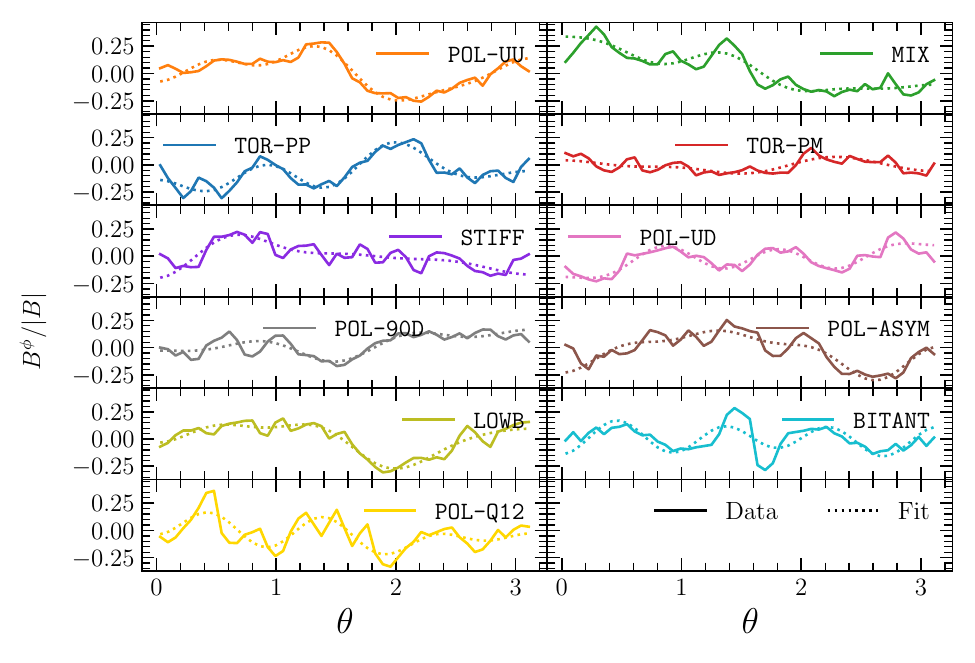}
    \caption{Mass averaged value of $B^\phi/|B|$ as a function of $\theta$, with truncated Legendre series fit. } 
 \label{fig:Bphi_theta}
\end{figure}

In Fig. \ref{fig:Bphi_theta} we demonstrate the mass averaged relative toroidal component of the 
magnetic field $B^\phi/|B|$ as a function of $\theta$ for the different configurations.
Generically across configurations we find a highly oscillatory field as a function of $\theta$, with 
the field oscillating between $\sim \pm 40\%$ of the field strength being contained in the toroidal 
component either co- or counter-aligned. We find the most extreme value for 
the toroidal field across our simulations in the polar regions, for the \MIXED~ configuration, which 
develops the strongest localised co-rotating toroidal field component directly above the merger 
remnant, at $\sim 42\%$ of the total field strength. The other, most strongly magnetised configuration \POLUD, in 
contrast, does not form such a coherent field in the ejecta. Immediately above and below the equatorial 
plane, we see a clear sign change in the toroidal field orientation in the benchmark configuration 
\POLUU, as well as \MIXED~ and \POLASYM, the 2 simulations most similar to this configuration; while 
in the \TORPP~ configuration we see the opposite behaviour, with the field counter rotating above the 
equatorial plane, and switching signs to become corotating below. The remaining configurations 
oscillate around zero.

Finally we comment on the multipolar structure of this field. We decompose the fraction of the
magnetic field in the toroidal component as a Legendre series in $\cos(\theta)$, $B^\phi / |B| = \sum_\ell a_\ell P_\ell(\cos\theta)$.
In Table \ref{tab:corr} we list the term in the series corresponding to the maximum coefficient, $\ell_\mathrm{max}$, and the 
associated correlation length of the magnetic field, given by the extraction radius over this value, $L = r_\mathrm{ex}/\ell_\mathrm{max}$.
In Fig. \ref{fig:Bphi_theta} we add the Legendre series fit, keeping terms up to $\ell=8$, which is larger than $\ell_\mathrm{max}$
for almost all configurations. The majority of the poloidally dominated runs have comparatively low values of $\ell_\mathrm{max}$, 
and are well described by the fit shown in Fig. \ref{fig:Bphi_theta}, 
 though we note that the enforced symmetry over the $z$ axis 
for the \BITANT~ configuration leads to a higher $\ell_\mathrm{max}$ than in the \POLUU~ case. The configurations with the highest 
$\ell_\mathrm{max}$, \STIFF, \TORPM~ and \POLQ~ show a distribution much closer to 0 at all $\theta$, and are dominated by small scale oscillations in the field
structure.

\begin{table*}[t]
  \centering    
  \begin{tabular}{c|ccccccccccc}        
    \hline
    Run & \POLUU & \MIXED & \TORPP & \TORPM & \BITANT & \LOWB & \POLUD & \POLLR & \POLASYM & \STIFF & \POLQ \\
\hline
    $\ell_\mathrm{max}$& 3 & 1 & 6 &17 & 7 & 3 &5  &4 &2 &17 & 19\\
    Correlation Length, $L$ (km)& 98.3 &295.0 &49.2 &17.4 &42.1 &98.3 &59.0 &73.8 &147.5 &17.4 & 15.5\\

    \hline
  \end{tabular}
  \caption{Correlation length for mass weighted toroidal field component}
 \label{tab:corr}
\end{table*}

\section{Conclusion}
\label{sec:con}

In this work we have performed a series of simulations of magnetised BNS mergers
with varying initial choices of magnetic field topology, also varying the 
EOS, mass ratio and symmetry assumptions of our numerical domain.
The impact of the choice of initial magnetic field topology has been found to have
a significant effect on the magnetic field evolution during the merger, as well
as in the ejected matter from the BNS merger. 

Firstly we have analysed the impact of different magnetic field configurations
on the GW signal of the BNS. The presence of a magnetic field
has minimal impact on the wave signal during the inspiral, but in the post-merger phase, 
different magnetic field configurations can have an impact on the evolution
of GW modes, compared with each other and with 
a reference unmagnetised configuration. We find that the dephasing 
between the dominant $(2,2)$ mode and leading subdominant $(2,1)$ and $(3,3)$ modes may be affected by different magnetic field configurations. This may lead to a 
change in the size of the dephasing that is tens of radians in magnitude, but
we note that this effect is highly sensitive to the reconstruction routine used and
that the overall magnitude of this effect seen in the $(3,3)$ mode is less than the
difference in dephasing that arises from changing reconstruction routine.

Subsequently we have investigated the impact of the 
initial field configuration on the magnetic field amplification. Here we have found a large impact, finding that the 
magnification arising during the KHI may be strongly amplified by the presence of 
anti-aligned magnetic fields, with anti-aligned poloidal and toroidal fields magnifying
considerably more than their aligned counterparts,
in agreement with local simulations. This feature may arise due to an increase in magnetic
tension in the anti-aligned case.
The amplification during this phase for toroidal fields however is strongly suppressed, due to the weaker magnetic field strengths in the region where the KHI is present.
This suggests that features associated to 
strong magnetic field development in the post-merger phase, such as the launching of 
a jet, may be delayed for models of neutron stars with 
strong toroidal fields, but that magnetic field amplification may be underestimated for generic configurations with different alignments of poloidal fields.

After the KHI has developed, we have also found that the magnetic field topology has a large
impact on the amplification due to winding, with the amplification slower in toroidally dominated configurations, but enhanced in the presence of asymmetries, and faster at late times for a combination of poloidal and toroidal fields that resemble the end state of the instabilities that a purely poloidal field will undergo. Enforcing a bitant symmetry also enhances this amplification. We find that, as a consequence, the Poynting flux from the merger remnant is enhanced for those configurations with larger and faster amplifications during the winding phase.

We have analysed the asymmetries that develop in our simulations when evolving symmetric initial data. We find that, independent of the specific topology of the initial data, that the 
size of the asymmetric magnetic field evolves at an approximately universal rate, saturating at approximately 10\% of the symmetric magnetic field in size. This suggests the presence of a spontaneous 
symmetry breaking bifurcation in the solution to the GRMHD equations, a feature that will be neglected when enforcing a bitant symmetry in such simulations.

We perform a detailed analysis of the ejected material from these simulations. For identical initial hydrodynamical configurations we find that the 
evolution of the hydrodynamics of the ejected material is broadly similar for different magnetic configurations. The nature of the magnetic fields
within the ejecta however can be strongly affected by the magnetic field configuration and amplification. 
We find that the magnetic field strength in the ejecta may vary by 
over an order of magnitude depending on the magnetic field initial data, with consequent
impacts on the $\beta^{-1}$ and magnetisation of the ejected material. 
We find the strongest magnetic fields in the ejecta at late times in the most strongly amplified configuration \POLUD, 
as well as the fastest growing \MIX~ configuration and the \POLQ~ configuration. These configurations most readily eject more highly magnetised material in the polar regions with comparatively larger magnetic fields up to $10^{11}G$, at densities below the maximum ejected density. 
We model the relationship between density and magnetic field strength, and density and $\beta^{-1}$ in the ejected material, and find a simple power law scaling that decays gradually over time, as more, less dense, highly magnetised material begins to be ejected in the polar regions. An enforced bitant symmetry forces a much weaker relationship between the density and magnetic field strength, than the equivalent full grid configuration, and leads to an enhanced field strength also in the ejecta. 

We have also analysed the magnetic field orientation in the ejected material in order to validate
existing models used to estimate thermalisation efficiencies of charged particles in the ejecta. We find that overall no dominant global field has formed in the ejecta by the end of our simulations,
and that consequently an approximately random model may provide a reasonable approximation to the 
field. This model increases the timescales over which charged particles may remain trapped in the magnetic field and given an opportunity to thermalise, impacting expected light curve luminosities in the kilonova, compared to a radial field. Towards the end of the timescales considered however we begin to see signs of a stronger toroidal field forming, especially for the \MIX~ configuration, in the polar region.

A fuller analysis of the properties of such configurations will require additional features not 
present in these simulations. Firstly the impact of neutrino effects must be included, through, 
for instance an M1 scheme \citep{Shibata:2011kx,Foucart:2015vpa,Radice:2021jtw}, in order to obtain 
a more accurate model of the ejecta temperature, entropy, and, key for nucleosynthesis estimates, the 
composition. Longer term simulations will be required to investigate the full impact of the magnetic field 
configuration on jet launching, and the use of subgrid modelling 
\citep{Radice:2017zta,Aguilera-Miret:2020dhz,Miravet-Tenes:2022ebf} may be key to accurately capturing the 
magnetic field in the remnant, and hence the later time magnetically driven ejecta. The impact of the magnetic 
field on the post-merger GW signal has been shown to be highly dependent on the 
choice of reconstruction method. Future simulations of these effects should probe even higher resolutions, to investigate 
the robustness of this feature, and its potential detectability.

\begin{acknowledgments}
  EG thanks Luciano Combi for helpful discussions.
  EG and DR acknowledge funding from the National Science Foundation under Grants No.~AST-2108467 and PHY-2407681.
  EG acknowledges funding from an Institute for Gravitation and Cosmology fellowship.
  DR acknowledges support from the Sloan foundation and from the National Science Foundation under Grant No.~PHY-2020275.
  JF and DR acknowledge U.S. Department of Energy, Office of Science, Division of Nuclear Physics under Award Number(s) DE-SC0021177 and DE-SC0024388.
  SB, BD, MJ acknowledge support by the EU Horizon under ERC Consolidator Grant,
  no. InspiReM-101043372.
  SB, MJ acknowledge support from the Deutsche Forschungsgemeinschaft (DFG) under MEMI (BE 6301/2-2, Projektnummer: 443239082) .
  PH acknowledges funding from the National Science Foundation under Grant No. PHY-2116686.
  The authors are indebted to A.~Celentano's PRİSENCÓLİNENSİNÁİNCIÚSOL.
  The numerical simulations were performed on TACC's Frontera (NSF LRAC
  allocation PHY23001) and on Perlmutter. This research used resources
  of the National Energy Research Scientific Computing Center, a DOE
  Office of Science User Facility supported by the Office of Science of
  the U.S.~Department of Energy under Contract No.~DE-AC02-05CH11231.
  Simulations were also performed on the national HPE Apollo Hawk
  at the High Performance Computing Center Stuttgart (HLRS).
  The authors acknowledge HLRS for funding this project by providing access
  to the supercomputer HPE Apollo Hawk under the grant numbers INTRHYGUE/44215
  and MAGNETIST/44288.
  Simulations were also performed on SuperMUC\_NG at the
  Leibniz-Rechenzentrum (LRZ) Munich.
  The authors acknowledge the Gauss Centre for Supercomputing
  e.V. (\url{www.gauss-centre.eu}) for funding this project by providing
  computing time on the GCS Supercomputer SuperMUC-NG at LRZ
  (allocations {\tt pn67xo}, {\tt pn76li}, {\tt pn68wi} and {\tt pn36jo}).
  Postprocessing and development run were performed on the ARA cluster
  at Friedrich Schiller University Jena.
  The ARA cluster is funded in part by DFG grants INST
  275/334-1 FUGG and INST 275/363-1 FUGG, and ERC Starting Grant, grant
  agreement no. BinGraSp-714626.
\end{acknowledgments}

\appendix

\section{Effect of Reconstruction Scheme}
\label{app:recon}

We have performed simulations with two different reconstruction schemes, PPM and WENOZ.
To benchmark the expected numerical errors introduced by varying the reconstruction
scheme, we perform the same run, \POLUU~ 
with both schemes. Here we demonstrate the 
 GW signal, magnetic field amplification and ejecta magnetisation for both runs
to quantify the variation due to reconstruction scheme.

\begin{figure}[t]
  \centering
    \includegraphics[width=0.49\textwidth]{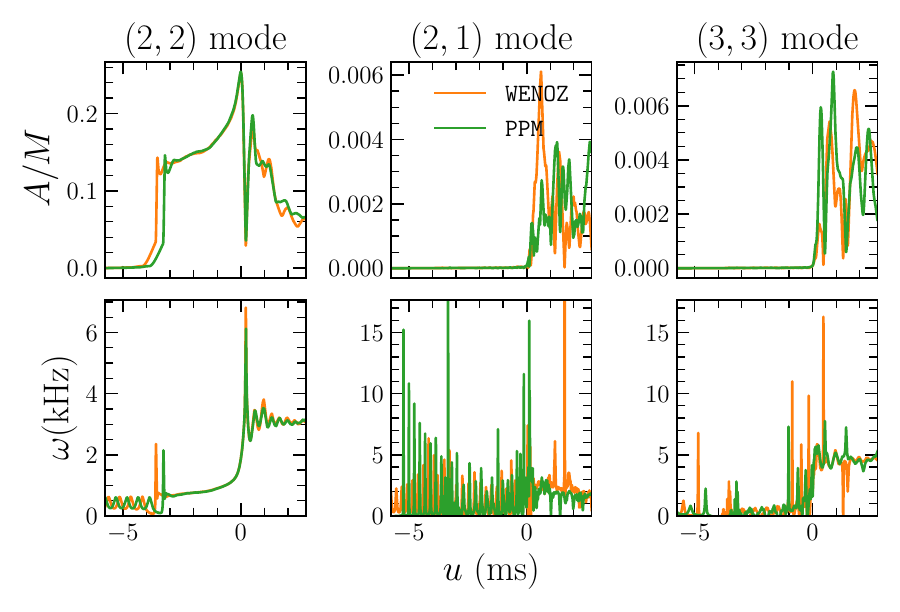}
    \caption{Upper row: The GW strain amplitude for configuration \POLUU~ with different reconstruction routines. Bottom row: The instantaneous GW frequency. Left, central and right columns show the $(\ell=2,m=2), (2,1), (3,3)$ modes respectively. }
 \label{fig:recon_GW}
\end{figure}
We first note that the merger time is, as expected, delayed for the WENOZ run
compared with the PPM run due to a reduced dissipation, with the merger occurring
$\sim 0.9$ms later in the WENOZ run. Accounting for the shifted time of merger,
we see in Fig. \ref{fig:recon_GW} the evolution of the (2,2) mode of the GW strain is almost identical 
up to merger in both amplitude and instantaneous frequency. In the post-merger phase,
we see small deviations in the (2,2) mode amplitude, and slightly larger oscillations
in the frequency for the WENOZ 
configuration. In the subdominant modes however, here
focussing on the (2,1) and (3,3) modes, we see a significant difference in the post-merger 
evolution. Immediately post-merger we see a large peak in the amplitude of the (2,1) mode
in the WENOZ configuration, a factor 2 larger than that of the PPM configuration, with later
oscillations in the PPM configuration larger than those in the WENOZ configuration. 
Conversely in the (3,3) mode, we see two
peaks in amplitude immediately post-merger for the PPM configuration. These peaks
are approximately coincident with, but larger than, corresponding peaks in the WENOZ configuration, 
with the initial peak a factor 4
larger in the PPM configuration. After these peaks we see the WENOZ amplitude growing above the PPM
values. For the instantaneous frequencies of the $(2,1)$ and $(3,3)$, the inspiral is dominated 
by noise, due to the overall small value of the amplitude of these modes, but in the post-merger,
other than isolated spikes in the frequency, the overall frequency evolution is broadly similar 
between the reconstruction methods.
To estimate the error associated to the
choice of reconstruction scheme on the GW dephasing discussed in Sec. \ref{ssec:gw}, we calculate the accrued dephasing $\Delta \phi_{21}$ and 
$\Delta \phi_{33}$ for the two \POLUU~ configurations. For the PPM (WENOZ) configuration, the
accrued dephasing is 29 (17) radians for the (2,1) mode and -30 (-48) radians for the (3,3) mode.
The gives an error of 12 and 18 radians respectively. These are large errors compared
to the maximal size of the dephasing effects seen in Sec. \ref{ssec:gw}, of size $39\%$ and $123\%$ respectively, though we note that it is still the case that for a consistent choice of reconstruction scheme,
a range of dephasings of size up to 12 and 8 radians can be found for the (2,1) and (3,3) modes respectively for different choices of magnetic field data.
The size of these errors emphasises the importance of robustly exploring the impact of not just resolution but also the choice of numerical scheme used to evolve the hydrodynamics, on any 
potentially detectable signal found in the post-merger GW.

\begin{figure}[t]
  \centering
    \includegraphics[width=0.49\textwidth]{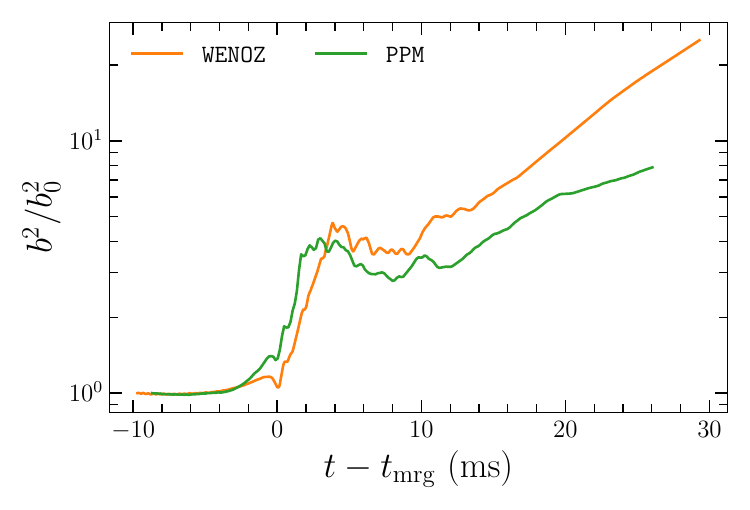}
    \caption{The magnetic field amplification for configuration \POLUU~ with different reconstruction routines.}
 \label{fig:recon_bamp}
\end{figure}
We now investigate the impact of reconstruction on the magnetic field amplification, demonstrated in Fig. \ref{fig:recon_bamp}. 
Again accounting for the different merger time, we find that the magnetic field amplification is relatively robust to the choice of reconstruction method. When drawing a comparison with the reconstruction
analysis of \cite{Cook:2023bag}, we see that the WENOZ method should preserve the central density of the
star more accurately than the PPM method, but the star surface may be preserved less well.
This may lead to a slightly more diffuse outer layer of the star, where we expect the KHI to 
develop. Therefore the development of this feature may be dependent on
the impact of the reconstruction scheme on the hydrodynamical evolution. We find that the PPM configuration shows an earlier start to the magnetic field amplification, 
and that the saturation of the KHI occurs more quickly than the corresponding WENOZ configuration, which we 
attribute to the merger process happening slightly quicker. Once the KHI has saturated, the amplification of
 the magnetic field between the two runs is very similar, with the WENOZ run amplifying by a factor of $\sim10\%$ 
more than the PPM run. After the saturation of the KHI, the intermediate phase progresses similarly for both configurations, 
with the magnetic field strength remaining constant, and lasting for approximately the same duration. As the winding phase 
begins, the slightly larger magnetic field for the WENOZ configuration results in a larger field generated by the end 
of the runs by a factor of $\sim{} 2$. We note that, while this is again a large factor if one wishes to make precise 
quantitative predictions from such simulations, a factor of 2 is subdominant to the two orders of magnitude seen across 
different magnetic field initial data configurations in Fig. \ref{fig:bamp}, and so we consider this analysis robust to choice of reconstruction method.

\begin{figure}[t]
  \centering
    \includegraphics[width=0.49\textwidth]{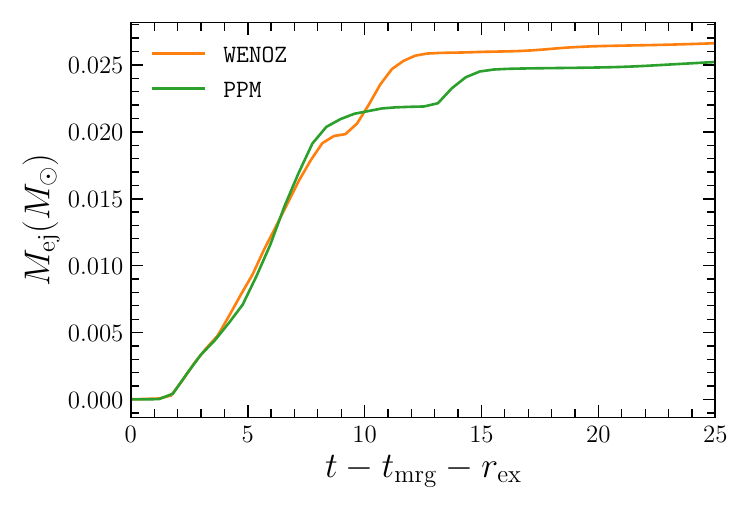}
    \caption{The ejected mass for configuration \POLUU~ with different reconstruction routines.}
 \label{fig:recon_ej_mass}
\end{figure}
Finally, we comment on the impact of reconstruction scheme on the ejecta, by comparing the total 
ejected mass, and the average magnetic field strength in the ejecta. In Fig. \ref{fig:recon_ej_mass},  for the PPM run we see that, 
initially after merger the first initial flux of ejecta is larger than the corresponding WENOZ run, 
leading to a larger mass ejection, but that subsequent ejections, corresponding to core bounces, 
are less extreme than in the WENOZ configuration, and consequently marginally more mass is ejected 
for the WENOZ run. At the end of the simulations the WENOZ run has ejected $\sim 0.027 M_\odot$, while
the PPM configuration ejects $\sim 0.025 M_\odot$, a $7\%$ error.

\begin{figure}
  \centering
    \includegraphics[width=0.49\textwidth]{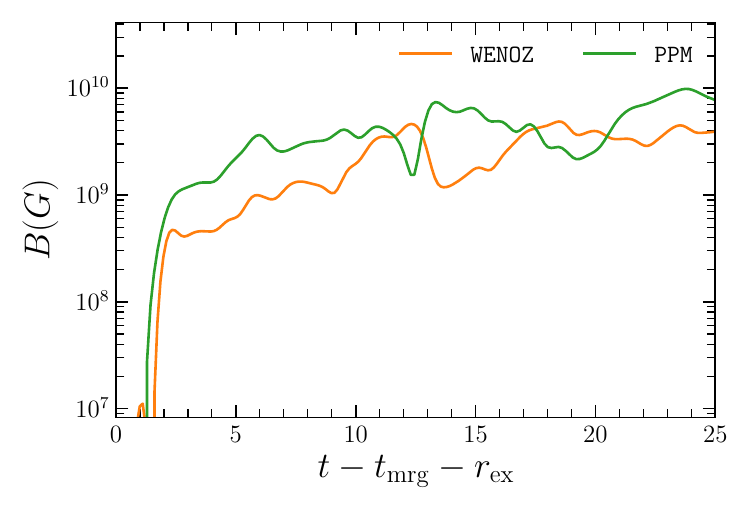}
    \caption{The magnetic field strength in the ejected material for configuration \POLUU~ with different reconstruction routines.}
 \label{fig:recon_ej_aveb}
\end{figure}

The magnetic field strength in the ejecta can also be affected by reconstruction scheme choice. In 
Fig. \ref{fig:recon_ej_aveb} we find that, contrary to the hierarchy during the winding phase, that the 
PPM configuration provides 
slightly more magnetised ejecta, at least at early times post-merger. At $\sim 5$ ms post-merger we 
see that the PPM configuration has on average a magnetic field $\sim{}4$ times stronger than in the 
WENOZ configuration. By the end of the simulations however, we see that the two configurations have 
magnetic fields of approximately equal magnitude on average, oscillating around a common value. 
The main impact on the ejecta from the reconstruction method appears to be on the unphysical low density 
material arriving at the extraction spheres before merger time. For the WENOZ configuration we see
more low density material arriving at early times, which appears to be absent for the PPM 
configurations, which again , supports the analysis of reconstruction methods in  \cite{Cook:2023bag}. 
These features can be seen in the low density tails of the 1D and 2D histograms in Sec. \ref{ssec:ej}, 
but we note that these tails contain very little mass, and that the bulk of the physical ejecta is 
comparatively unaffected, as justified by the comparatively small errors found in this direct comparison.

\end{document}